\documentclass[twocolumn,twoside]{IEEEtran}

\usepackage{xcolor,soul,framed} %,caption

\colorlet{shadecolor}{yellow}
\usepackage[pdftex]{graphicx}
\graphicspath{{../pdf/}{../jpeg/}}
\DeclareGraphicsExtensions{.pdf,.jpeg,.png}

\usepackage[cmex10]{amsmath}
\usepackage{array}
\usepackage{mdwmath}
\usepackage{mdwtab}
\usepackage{eqparbox}
\usepackage{url}
\usepackage{flushend,cuted}
\usepackage[noend]{algpseudocode}
\usepackage{algorithmicx,algorithm}
\usepackage{subfigure}
\usepackage{caption}
\usepackage{cite}
\usepackage{xcolor}
\usepackage{epstopdf}
\usepackage{cases}
\usepackage{amscd,textcomp,amssymb,epsfig,graphics,epsf,color,amsmath,balance,cite}
\usepackage{multirow}
\usepackage{booktabs}
\usepackage{stfloats}
\usepackage{setspace}
\usepackage{graphicx}
\usepackage{amsmath, amsthm, amsfonts, amssymb, amsbsy}
\usepackage{dsfont}
\usepackage{color}
\usepackage{bm}

\allowdisplaybreaks[4]%åè®žå¬åŒè·šé¡µ

\begin{document}
%Integrated
\title{Joint OAM Radar-Communication Systems:\\ Target Recognition and Beam Optimization}%
%, Performance Analysis
\author{Wen-Xuan Long,~\IEEEmembership{Graduate Student Member,~IEEE,}~Rui Chen,~\IEEEmembership{Member,~IEEE,}\\Marco Moretti,~\IEEEmembership{Member,~IEEE,} Wei Zhang,~\IEEEmembership{Fellow,~IEEE,} and Jiandong Li,~\IEEEmembership{Fellow,~IEEE}
%Wei Zhang,~\IEEEmembership{Fellow,~IEEE}

\thanks{W.-X. Long is with the State Key Laboratory of Integrated Service Networks (ISN), Xidian University, Shaanxi 710071, China, and also with the University of Pisa, Dipartimento di Ingegneria dell'Informazione, Italy (e-mail: wxlong@stu.xidian.edu.cn).}
\thanks{R. Chen is with the State Key Laboratory of Integrated Service Networks (ISN), Xidian University, Shaanxi 710071, China, and also with the National Mobile Communications Research Laboratory, Southeast University, Nanjing 210018, China (e-mail: rchen@xidian.edu.cn).}
\thanks{M. Moretti is with the University of Pisa, Dipartimento di Ingegneria dell'Informazione, Italy (e-mail: marco.moretti@iet.unipi.it).}
\thanks{W. Zhang is with the School of Electrical Engineering and Telecommunications, The University of New South Wales, Australia, (e-mail: w.zhang@unsw.edu.au).}
\thanks{J. Li is with the State Key Laboratory of Integrated Service Networks (ISN), Xidian University, Shaanxi 710071, China (e-mail: jdli@mail.xidian.edu.cn).}
}

\maketitle

\thispagestyle{empty}
% ==================================================
\begin{abstract}
Orbital angular momentum (OAM) radars are able to estimate the azimuth angle and the rotation velocity of multiple targets without relative motion or beam scanning. Moreover, OAM wireless communications can achieve high spectral efficiency (SE) by utilizing a set of information-bearing modes on the same frequency channel. Benefitting from the above advantages, in this paper, we design a novel radar-centric joint OAM  radar-communication (RadCom) scheme based on uniform circular arrays (UCAs), which modulates information signals on the existing OAM radar waveform. % to realize the win-win result between the OAM radar and OAM communications. For the considered UCA-based joint OAM RadCom system,
In details, we first propose an OAM-based three-dimensional (3-D) super-resolution position estimation and rotation velocity detection method, which can accurately estimate the 3-D position and rotation velocity of multiple targets. % by taking full advantage of the phase structure of OAM beams.
Then, we derive the posterior Cram\'{e}r-Rao bound (PCRB) of the OAM-based estimates and, finally, we analyze the transmission rate of the integrated communication system. To achieve the best trade-off between  imaging and communication, the transmitted integrated OAM beams are optimized by means of an exhaustive search method. %Due to the complexity of the optimization problem, we apply an exhaustive search method to solve the problem.
Both mathematical analysis and simulation results show that the proposed radar-centric joint OAM RadCom scheme can accurately estimate the 3-D position and rotation velocity of multiple targets while ensuring the transmission rate of the communication receiver, which can be regarded as an effective supplement to  existing joint RadCom schemes.
\end{abstract}

\begin{IEEEkeywords}
Orbital angular momentum (OAM), joint radar-communication, target recognition, posterior Cram\'{e}r-Rao bound (PCRB), beam optimization, uniform circular array (UCA).
\end{IEEEkeywords}

\section{Introduction}

Novel service requirements are the driving force behind the evolution of wireless communication networks. The rapid development of emerging applications, such as holographic video, digital twin, virtual reality and auto-pilot driving, results in a neverending growth in mobile data traffic. It is reported that the capacity of next generation wireless communication networks will reach $100$ times that of the existing 5G networks \cite{Zhang20196G,Long2021A}. To meet these requirement, more and more high frequency bands such as millimeter-wave and terahertz bands are being licensed \cite{WRC}. Unfortunately, this further leads to increased congestion of the frequency spectrum where existing radar and other sensing systems reside. To solve this problem, the concept of a novel joint RadCom system is proposed, in which the previously competing radar sensing and communication operations can be implemented simultaneously by sharing a single hardware platform and a joint signal processing framework. Based on design priorities and the underlying signal, existing joint RadCom schemes can be classified into communication-centric schemes, radar-centric schemes and  joint design schemes. Due to the potential close cooperation between radar sensing and wireless communications,  joint RadCom schemes are recognized as a key approach in significantly improving spectrum efficiency, reducing device size, cost and power consumption \cite{zhang2021overview}.

In recent years, significant progress in the research of joint RadCom systems has been made \cite{Zheng2019Radar,Feng2020Joint,zhang2021overview}. In \cite{Mealey1963A}, the concept of joint RadCom systems was proposed for the first time, providing new ideas for the research of radar and communication technologies. In \cite{Quan2014Radar}, several joint sharing schemes, such as time-sharing, frequency-sharing and beam-sharing, are  discussed in details, laying the foundation for the design of joint RadCom systems in different scenarios. Thereafter, a variety of emerging integrated beam design schemes have been studied. In \cite{Liu2018Toward}, several optimization-based waveform designs are proposed for a joint multiple-input multiple-output (MIMO) RadCom system. % to realize a trade-off between imaging and communication performances in different scenarios.
In \cite{Zheng2018On}, an orthogonal frequency division multiplexing (OFDM)-based scheme is applied to a MIMO RadCom system, which can achieve two-dimensional (2-D) position estimation and velocity detection of targets while communicating. In \cite{rihan2018non}, a novel three-phases non-orthogonal multiple access (NOMA)-based spectrum sharing strategy is applied to a joint MIMO RadCom system. In \cite{Ahmed2019Distributed}, a novel power distribution strategy is applied to a distributed MIMO RadCom system to improve the communication data rate while ensuring good radar imaging performance. In \cite{Keskin2021MIMO}, a novel inter-carrier interference (ICI)-aware sensing algorithm is applied to a MIMO-OFDM RadCom system to estimate delay-Doppler-angle parameters of multiple targets in high-mobility scenarios. In \cite{Zhou2021Performance}, a %joint-optimized
phase-modulated waveform is applied to a MIMO RadCom system to find the best trade-off between  radar and communication performance.

However, there are still several technical challenges for the design scheme of a RadCom system. \emph{One such problem is that  in a RadCom system the azimuth angle and the rotation velocity of the targets cannot be obtained unless there is  a relative motion between the dual-function transceiver and the targets.}
%\emph{A specific problem is that limited by the transmitted integrated waveform, the azimuth angle and rotation velocity of targets cannot be obtained by the dual-function transceiver of the existing joint RadCom system when is no relative motion between the dual-function transceiver and targets, which is not conducive to the position estimation and status recognition of targets.}
One promising approach to solve this problem is to employ integrated OAM waveforms. The phase front of an electromagnetic (EM) wave carrying OAM rotates with azimuth exhibiting a helical structure $e^{j\ell\phi'}$ in space \cite{Allen1992Orbital}, where $\phi'$ is the transverse azimuth and $\ell$ is an unbounded integer defined as \emph{OAM topological charge} or \emph{OAM mode number}. This helical phase structure provides a new degree of freedom for radar application and information transmission.

For radar sensing, the vortex EM wave carrying OAM can be seen as multiple plane EM waves that simultaneously illuminate the target from continuous azimuth, achieving thus angular diversity without relative motion or beam scanning. Therefore, OAM-based schemes are regarded as a promising approach to provide the azimuthal resolution without relative motion \cite{Guo2013Electromagnetic,lin2016improved,Chen2018OAMradar,Liu2020Microwave,wang2021Object,Yuan2021Resolution}.
%In \cite{Guo2013Electromagnetic}, an OAM-based azimuth angle detection method is proposed for the first time, which opens a new perspective for existing radar sensing techniques. Then, a variety of high-resolution methods are applied to the OAM-based radar system to further improve the azimuth resolution \cite{lin2016improved, Chen2018OAMradar,Liu2020Microwave,wang2021Object,Yuan2021Resolution}.
In particular, when vortex EM waves illuminate spinning targets, the rotational Doppler shift caused by spinning targets is proportional to the spinning velocity and OAM mode \cite{Lavery2013Detection}. Based on this property,  OAM-based radars can also be used to estimate the spinning velocity of targets even when there is no radial motion between the radar and targets \cite{Courtial1998Measurement,Zhao2016Measurement,Liu2017Microwave}. In \cite{Courtial1998Measurement}, the rotational Doppler shift of spinning targets is first observed. In \cite{Zhao2016Measurement}, a method based on the phase structure of OAM beams is proposed to estimate the spinning velocity of the targets. In \cite{Liu2017Microwave},  both simulations and proof-of-concept experiments show the effectiveness of the OAM radar for imaging of spinning targets.

\vspace{0.0cm}
\begin{figure}[t] %figure1
\setlength{\abovecaptionskip}{-0.0cm}   %è°æŽåŸçæ é¢äžåŸè·çŠ»
\setlength{\belowcaptionskip}{-0.4cm}   %è°æŽåŸçæ é¢äžäžæè·çŠ»
\footnotesize
\begin{center}
\includegraphics[width=8.9cm,height=6.0cm]{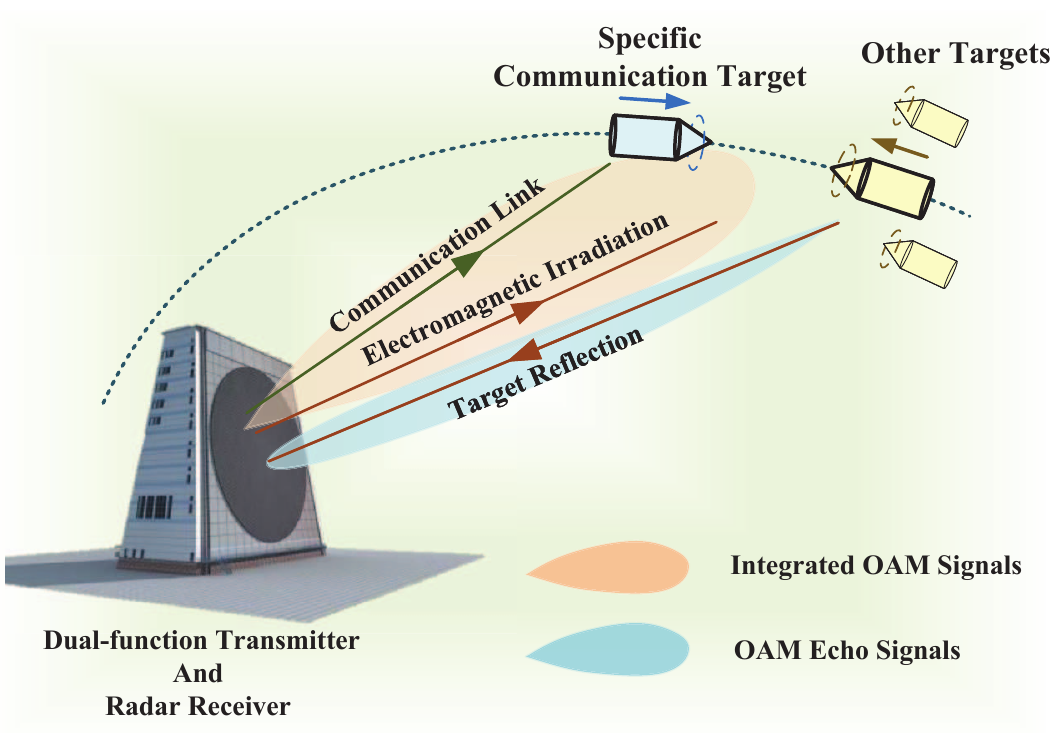}
\end{center}
\caption{An application scenario \cite{Cole2010Missile} of the proposed radar-centric joint OAM RadCom system.}
\label{Fig1}
\end{figure}

Furthermore, for wireless communications, OAM-based schemes enable a novel coaxial multiplexing approach, which utilizes a set of information-bearing modes on the same frequency channel to achieve high spectral efficiency (SE) \cite{Chen2020Orbital,Tamburini2012Encoding,Mahmouli20134,Yan2014High,Ren2017Line,Chen2018A,Chen2020Multi,Long2021Joint,Long2021AoA,Yagi2021200}. In \cite{Tamburini2012Encoding}, a  wireless transmission experiment based on OAM is realized for the first time by successfully multiplexing two different radio signals at the same frequency. Moreover, a 4 Gbps uncompressed video transmission link over a 60 GHz OAM radio channel is implemented in \cite{Mahmouli20134}.
%It is shown in \cite{Yan2014High} that multiplexing four dual-polarized OAM modes (mode number $\ell=\pm1,\pm3$) generated by spiral phase plates (SPPs) can achieve a 32 Gbit/s data rate in a wireless communication link at 28 GHz.
In \cite{Ren2017Line}, a $2\times2$ antenna aperture architecture, where each aperture multiplexes two OAM modes, is implemented in the 28 GHz band achieving a 16 Gbit/s transmission rate. In \cite{Yagi2021200}, a communication link with a  transmission rate of over 200 Gbit/s is obtained  by multiplexing five  dual-polarized OAM modes (mode number $\ell=0, \pm1,\pm2$) in the 28 GHz band.

% Considering the advantages of OAM technology in radar sensing and wireless communications, we propose a novel integrated OAM RadCom scheme, including the three-dimensional (3-D) position estimation, rotation velocity detection and specific node communication, which is a promising approach for realizing spectrum sharing of wireless communication services and radar sensing services in the next 6G network. The novelty and major contributions of this paper are
Up to now, OAM-based schemes have been widely studied in both radar sensing and wireless communications. However, little has been done about a joint OAM RadCom technology. Considering the problems faced by the existing joint RadCom schemes and the significant advantages of OAM radar sensing, we propose a novel radar-centric joint OAM RadCom scheme that includes  three-dimensional (3-D) super-resolution position estimation, rotation velocity detection and specific target communication. The novelty and major contributions of this paper are summarized as follows:
\begin{enumerate}
\item
We present a novel radar-centric joint OAM RadCom scheme based on uniform circular arrays (UCAs), which modulates information signals over the existing OAM radar waveform. By taking full advantage of the phase structure of OAM beams, the proposed  OAM RadCom scheme can effectively estimate the azimuth angle and rotation velocity of targets without beam scanning, while at the same time communicating with the target.
\item
We discuss a novel OAM-based multi-target imaging method, which includes  3-D super-resolution position estimation and rotation velocity detection. By using the super-resolution multiple signal classification (MUSIC) algorithm in the frequency and OAM mode domain, the proposed method solves the problem of the restricted elevation resolution faced by  existing OAM-based imaging methods. Moreover, the proposed method can estimate the rotation velocity of multiple targets simultaneously by taking full advantage of the rotational Doppler characteristics of OAM echo signals, providing a basis for distinguishing the type of targets.
%the 3-D position of multiple targets can be accurately estimated. Second, an OAM-based 3-D super-resolution position estimation method is proposed. {\color{red}By using the super-resolution multiple signal classification (MUSIC) algorithm for OAM echo signals in the frequency domain and OAM mode domain, the proposed method solves the problem of the restricted elevation resolution faced by the existing OAM-based target imaging, and realizes the 3-D super-resolution imaging of multiple targets.} Besides, an OAM-based rotation velocity detection method is proposed, which takes full advantage of the rotational Doppler characteristics of OAM echo signals to estimate the rotation velocity of multiple targets.
\item
The imaging and communication performances of the proposed OAM-based RadCom system are analyzed by deriving for the first time the posterior Cram\'{e}r-Rao bound (PCRB) of the OAM-based target imaging. Thereafter, the transmitted integrated OAM beams are optimized by minimizing the PCRB under the constraints dictated by the requirements of the communication part.
\end{enumerate}
%äžç»Žäœçœ®äŒ°è®¡éšåçåæ°æ§ïŒæä»¬åæ°å°å°ç»åžçMUSICç®æ³åæ¶åºçšäºé¢ååæš¡æåïŒå®ç°äºQäžªèç¹çè¶åèŸšäžç»Žäœçœ®äŒ°è®¡ã

The remainder of this paper is organized as follows. Based on the feasibility of generating and receiving OAM beams by UCAs \cite{Liu2016Generation,Chen2018A}, we model the UCA-based radar-centric joint OAM RadCom system in Section II. In Section III, the OAM-based 3-D position estimation and rotation velocity detection methods are proposed. After that, we analyze the PCRB and the data rate of the joint OAM RadCom system, and optimize the transmitted integrated OAM beams in Section IV. Simulation results are shown in Section V and conclusions are summarized in Section VI.

{\sl Notations}: Unless otherwise specified, matrices are denoted by bold uppercase letters (i.e., $\mathbf{A}$), vectors are represented by bold lowercase letters (i.e., $\mathbf{a}$), and scalars are denoted by normal font (i.e., $a$). $(\cdot)^{\mathrm{T}}$, $(\cdot)^{\mathrm{H}}$, $(\cdot)^{-1}$ and $(\cdot)^{\dagger}$ stand for the transpose, Hermitian transpose, inverse and Moore-Penrose pseudo-inverse of the matrices. $|\cdot|$ and $\textrm{Re}$$[$ $\cdot$ $]$ stand for the modulus and the real part of the complex numbers. $\mathbb{E}\{\cdot\}$ denotes the statistical expectation.
%, $\lfloor$ $\cdot$ $\rfloor$ denotes the floor function

\vspace{0.0cm}
\section{Joint OAM Radar-Communication System}
\begin{figure*}[t]
\setlength{\abovecaptionskip}{-0.1cm}   %è°æŽåŸçæ é¢äžåŸè·çŠ»
\setlength{\belowcaptionskip}{-0.4cm}   %è°æŽåŸçæ é¢äžäžæè·çŠ»
\begin{center}
\includegraphics[scale=0.405]{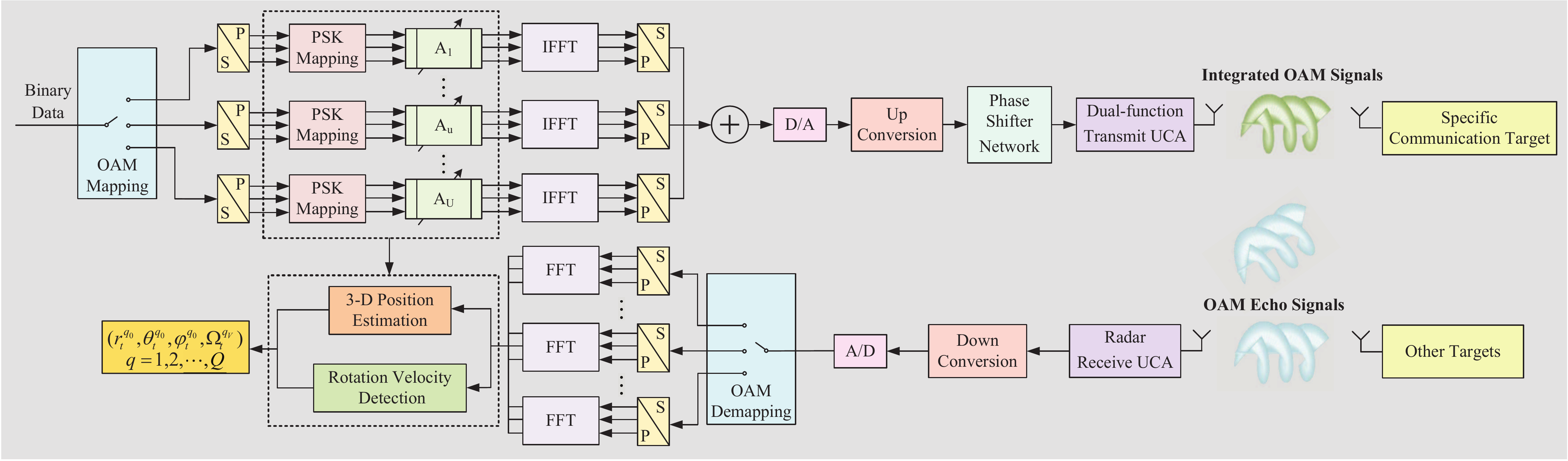}
\end{center}
\caption{The block diagram of the radar-centric joint OAM RadCom system.}
\label{Fig2}
\end{figure*}

\vspace{0.0cm}
\subsection{System Model}
\vspace{0.0cm}
%Further, these $Q$ nodes include cooperative nodes that can communicate with the dual-function transmitter and non-cooperative nodes that cannot communicate with the dual-function transmitter
Employing UCA is a popular way to generate and receive radio OAM beams due to its simple structure and the multi-mode multiplexing ability \cite{Liu2015Orbital,Liu2016Generation,Chen2020Multi,Long2021Joint}. Thus, we consider a UCA-based radar-centric joint OAM RadCom system as shown in Fig.\ref{Fig1}, which consists of one dual-function transmitter with an $M$-element UCA, one radar receiver with an $N$-element UCA and $Q$ targets with a single communication antenna. To simplify the mathematical descriptions that follow, we assume that the dual-function transmit and radar receive UCAs are co-located and have the same radius so that a target in the far-field is at the same spatial angle with respect to both the transmitter and receiver.

In the proposed joint OAM RadCom system, the dual-function transmitter periodically transmits  integrated OAM beams modulated with information. For easier analysis, we suppose that the transmitter sends data only to a specific target in each period. Therefore, by transmitting integrated OAM signals, information can be transmitted to the specific target, and the state parameters of all the targets can be estimated by OAM echo signals at the radar receiver. Moreover, we assume that the wireless communication between the dual-function transmitter and the specific target is performed through a line of sight (LoS) transmission and the communication channel can be accurately estimated \cite{Chen2020Multi,Long2021Joint,Long2021AoA}.%åŒåºåŸ1çåºæ¯

\subsection{Signal Model}
%transmit waveform
To provide the range-angle-dependent beam pattern for distinguishing multiple targets, we consider an OAM-OFDM-based signal scheme as shown in Fig.\ref{Fig2}. In the proposed radar-centric joint OAM RadCom system, we assume that the dual-function transmitter sends integrated signals by $U$ OAM modes $(U$$\leq$$M)$ and $W$ subcarriers. To avoid inter-mode interference, we suppose that the $U$ integrated OAM signals are transmitted sequentially in a specific sequence during each period \cite{lin2016improved,Chen2018OAMradar,Liu2020Microwave}. Meanwhile, to embed communication information into integrated OAM beams, we introduce $M_p$-ary PSK in each transmitted OAM waveform \cite{Bekar2021Joint}. Moreover, considering the different gains of different mode integrated OAM beams, we design a weight factor $A_u$ for the power distribution among the different mode OAM beams to optimize the performances of the system. Then, the $u$-th signals transmitted by the $m$-th transmit element at the $w$-th subcarrier can be expressed as
%{\color{red}To generate the joint OAM beams whose phase transforms uniformly with azimuth, the traditional spatial-domain beamforming is not suitable for the proposed OAM JRC system. Hence, we consider a time-domain beamforming scheme \cite{Mucci1984comparison,Hamid2014Performance} for the transmitted joint OAM beams. Meanwhlie,} to embed communication information into {\color{red}joint OAM waveform}, we introduce $M_p$-ary PSK in each {\color{red}transmitted OAM waveform}. Moreover, to avoid inter-mode interference, we suppose that {\color{red} the $U$ joint OAM symbols} are transmitted sequentially in a specific sequence during each period \cite{Liang2018Mode}, and the $u$-th signals transmitted by the $m$-th transmit element at the $w$-th subcarrier can be expressed as
%
\begin{align}
x_m(\ell_u,k_w)=A_u e^{i\ell_u\varphi_m}s(\ell_u,k_w), m=1,2,\cdots,M,
\end{align}
where $i$ is the imaginary unit, $\varphi_m=2\pi(m-1)/M$ is the azimuthal angle of the $m$-th transmit element, $s(\ell_u,k_w)=e^{i\phi_p}$ is the $u$-th integrated OAM symbol at the $w$-th subcarrier, $\phi_p\in [0,\frac{2\pi}{M_p},\cdots, \frac{2\pi(m_p-1)}{M_p}]$, $k_w=2\pi f_w/c$ is the wave number corresponding to subcarrier frequency $f_w$, $c$ is the speed of light in vacuum, $A_u$ is the weight of the $u$-th integrated OAM symbol with $\sum_{u=1}^{U}|A_u|^2=1$.

\vspace{0.0cm}
\subsubsection{OAM Radar Echo Signal Model}

For an arbitrary point $P'(r_t,\theta_t,\varphi_t)$ in the far field at arbitrary time $t$, the electric field intensity $E_T(\mathbf{r}_t,\ell_u,k_w)$ generated by the dual-function transmitter can be written as \cite{Guo2013Electromagnetic,Long2021AoA}
\begin{align} \label{Et}
&E_T(\mathbf{r}_t,\ell_u,k_w)=\nonumber\\
&-j\frac{\mu_0\omega_w}{4\pi}\sum_{m=1}^{M}x_m(\ell_u,k_w)
\int|\mathbf{r}_t-\mathbf{r}_m|^{-1}e^{ik_w|\mathbf{r}_t-\mathbf{r}_m|}dV_m\nonumber\\
&\overset{(a)}\approx-j\frac{A_u\mu_0\omega_w d}{4\pi}\frac{e^{ik_wr_t}}{r_t}\sum_{m=1}^{M}e^{-i(\mathbf{k}_{w,t}\cdot\mathbf{r}_m-\ell_u\varphi_m)}s(\ell_u,k_w)\nonumber\\
&\overset{(b)}\approx\!\!-j\frac{A_u\mu_0\omega_w dMe^{ik_wr_t}e^{i\ell_u\varphi_t}}{4\pi r_t}i^{-\ell_u}{J_{\ell_u}}(k_wR\sin\theta_t)s(\ell_u,k_w),
\end{align}
where $\mathbf{r}_t$ is the position vector of $P'(r_t,\theta_t,\varphi_t)$, $j$ is the current density of the dipole, $\mu_{0}$ is the magnetic conductivity in the vacuum, $\omega_w=2\pi f_w$ is the circular frequency, $d$ is the electric dipole length, $\int(\cdot)dV_m$ is the integral for the dipole in the $m$-th element of the dual-function transmit UCA, $J_{\ell_u}(\cdot)$ is the ${\ell_u}$th-order Bessel function of the first kind, $R$ is the radius of UCAs. In \eqref{Et}, (a) applies the approximation $|\mathbf{r}_t-\mathbf{r}_m|\approx r_t$ for amplitudes and $|\mathbf{r}_t-\mathbf{r}_m|\approx r_t-\mathbf{\hat{r}}_t\cdot\mathbf{r}_m$ for phases in the far field, where $\mathbf{\hat{r}}_t$ is the unit vector of $\mathbf{r}_t$, $\mathbf{r}_m=R\left(\mathbf{x'}\cos\varphi_m+\mathbf{y'}\sin\varphi_m\right)$, $\mathbf{x'}$ and $\mathbf{y'}$ are the unit vectors of X-axis and Y-axis of the coordinate system at the transmit/receive UCA, respectively, and (b) holds when $M$ is large enough.

We assume that the radar receiver continually receives OAM echo signals, even when the dual-function transmitter is transmitting integrated OAM signals. Although partially integrated OAM signals in this case will leak to the radar receiver, the transmission leakage can be restrained by transmission leakage cancellation techniques, such as antenna isolation and adaptive interference cancellation. Therefore, in this paper, the transmission leakage is ignored.

Assuming that the $q$-th target is discretized into $P$ ideal scattering points, the baseband signals received by the $n$-th element of the radar receive UCA can be written as
\begin{align}\label{En}
&E^n_R(\ell_u,k_w)=\sum_{q=1}^{Q}\sum_{i=1}^{P}E_T(\mathbf{r}^{q_{i}}_t,\ell_u,k_w)\sigma_{q_{i}}\cdot\nonumber\\
&\quad \quad\quad\quad\quad\ \int|\mathbf{r}^{q_{i}}_t-\mathbf{r}_n|^{-1}e^{ik_w|\mathbf{r}^{q_{i}}_t-\mathbf{r}_n|}dV_n+n(\ell_u,k_w)\nonumber\\
&\!\approx\!\sum_{q=1}^{Q}\!\sum_{i=1}^{P}E_T(\mathbf{r}^{q_{i}}_t,\ell_u,k_w)\!\cdot\!\!\big[\sigma_{q_{i}} d\frac{e^{ik_wr^{q_{i}}_t}}{r^{q_{i}}_t}e^{-i\mathbf{k}_{w,t}^{q_{i}}\mathbf{r}_n}\big]\!+\!n(\ell_u,k_w),
\end{align}
where
%the $q$-th target is discretized into $P$ ideal scattering points,
$\mathbf{r}^{q_{i}}_t$ and $\sigma_{q_{i}}$ are the position vector and the radar cross section (RCS) of the $i$-th scattering point in the $q$-th target, $\mathbf{r}_n=R\left(\mathbf{x'}\cos\varphi_n+\mathbf{y'}\sin\varphi_n\right)$, $\varphi_n=2\pi(n-1)/N$ is the azimuthal angle of the $n$-th receive element, $\int(\cdot)dV_n$ is the integral for the dipole in the $n$-th element of the radar receive UCA, $n(\ell_u,k_w)$ is additive white Gaussian noise (AWGN) with zero mean and variance $\xi^2$.

All the signals received by the $N$ elements of the radar receiver are combined, and the OAM echo signal on the $u$-th mode at the $w$-th subcarrier can be expressed as
\begin{align} \label{Er}
&E_R(\ell_u,k_w)=\sum_{n=1}^{N}E^n_R(\ell_u,k_w)+n(\ell_u,k_w)\nonumber\\
&\approx\!-j\frac{A_u\mu_0\omega_w d^2MN}{4\pi}i^{-\ell_u}\sum_{q=1}^{Q}\sum_{i=1}^{P}\sigma_{q_{i}} \frac{e^{i2k_wr^{q_{i}}_t}}{{(r^{q_{i}}_t)}^2}e^{i\ell_u\varphi^{q_{i}}_t}\times\nonumber\\
&\quad {J_{\ell_u}}(k_wR\sin\theta^{q_{i}}_t){J_0}(k_wR\sin\theta^{q_{i}}_t)s(\ell_u,k_w)+n(\ell_u,k_w),
\end{align}
where $(r^{q_{i}}_t,\theta^{q_{i}}_t,\varphi^{q_{i}}_t)$ is the coordinate of the $i$-th scattering point of the $q$-th target in the coordinate system of the transmit/receive UCA at the time instant $t$.

\vspace{0.0cm}
\subsubsection{Specific Target Communication Signal Model}
For the specific target whose centroid is located at $(r^{\bar{q}_{0}}_t,$$\theta^{\bar{q}_{0}}_t,$$\varphi^{\bar{q}_{0}}_t)$, the received equivalent baseband signal on the $u$-th OAM mode at the $w$-th subcarrier can be expressed as
\begin{align} \label{y}
&y_{\bar{q}}(\ell_u,k_w)=\sum_{m=1}^{M}h^{\bar{q}}_m(k_w)x_m(\ell_u,k_w)+z_{\bar{q}}(\ell_u,k_w)\nonumber\\
&\!\!\approx\frac{A_u\beta}{2k_wr^{\bar{q}_{0}}_t}\exp\big(-ik_wr^{\bar{q}_{0}}_t\big)\sum_{m=1}^{M}\exp\big[i\ell_u\varphi_m-\nonumber\\
&\quad ik_wR\sin\theta^{\bar{q}_{0}}_t\cos(\varphi^{\bar{q}_{0}}_t+\varphi_m)\big]s(\ell_u,k_w)+z_{\bar{q}}(\ell_u,k_w),
\end{align}
where
\begin{align*}
&h^{\bar{q}}_m(k_w)\approx\nonumber\\
&\!\!\frac{\beta}{2k_wr^{\bar{q}_{0}}_t}\exp\big[-ik_wr^{\bar{q}_{0}}_t-ik_wR\sin\theta^{\bar{q}_{0}}_t\cos(\varphi^{\bar{q}_{0}}_t+\varphi_m)\big]
\end{align*}
is the channel coefficient from the $m$-th element of the transmit UCA to the receive antenna of the specific target, $\beta$ models all constants relative to the antenna elements and their patterns, and $z_{\bar{q}}(\ell_u,k_w)$ is the AWGN with zero mean and variance $\xi^2$.

\section{OAM-based Rotating Target Imaging}
In this section, we propose a novel OAM-based multi-target imaging method, including  3-D super-resolution position estimation and rotation velocity detection. The proposed method breaks through the limitation of elevation resolution of the existing OAM-based imaging method while providing a basis for distinguishing the type of targets.
%{\color{red}In this section, we first propose an OAM-based 3-D super-resolution position estimation method, which breaks through the limitation of the elevation resolution of the existing OAM-based imaging method and realizes the 3-D super-resolution imaging of multiple targets. Then, we propose an OAM-based rotation velocity detection method, which takes full advantage of the rotational Doppler characteristics of OAM echo signals to estimate the rotation velocity of multiple targets, providing a basis for distinguishing the type of targets.}

%In this section, we propose a novel OAM-based multi-target imaging method, including the OAM-based 3-D super-resolution position estimation and rotation velocity detection, which breaks through the limitation of elevation resolution of the existing OAM-based imaging method while providing a basis for distinguishing the type of targets.
%which can effectively distinguish the type of nodes according to the rotation velocity of nodes while accurately estimating 3-D position of all nodes
%three-dimensional (3-D)
\subsection{OAM-based 3-D Position Estimation}

\subsubsection{Problem Formulation}

After the radar receiver is bit synchronized, the OAM echo signal on the $u$-th mode at the $w$-th subcarrier in \eqref{Er} can be simplified as
\begin{align}
&E'_R(\ell_u,k_w)=-\frac{E_R(\ell_u,k_w)}{|\eta(\ell_{u},k_w)|}\cdot \frac{s(\ell_{u},k_w)^*}{|s(\ell_{u},k_w)|}\cdot i^{\ell_u}\nonumber\\
&\!=\sum_{q=1}^{Q}\!\sum_{i=1}^{P}\!\sigma_{q_{i}} \frac{e^{i2k_wr^{q_{i}}_t}}{{(r^{q_{i}}_t)}^2}e^{i\ell_u\varphi^{q_{i}}_t}{J_{\ell_u}}(k_wR\sin\theta^{q_{i}}_t){J_0}(k_wR\sin\theta^{q_{i}}_t)\nonumber\\
&+n'(\ell_u,k_w),
\label{ER2}
\end{align}
where $\eta(\ell_{u},k_w)=-j\frac{A_u\mu_0\omega_w d^2MN}{4\pi}s(\ell_{u},k_w)$ consists of the known parameters of the radar receiver, $n'(\ell_u,k_w)$ is the noise. Then, all the signals received on the $U$ OAM modes at the $W$ subcarriers can be collected in the matrix
\begin{equation}\label{matrixE}
\mathbf{E}'_R\!=\!
\begin{bmatrix}
E'_R(\ell_1,k_1)             & E'_R(\ell_1,k_2)      & \cdots    & E'_R(\ell_1,k_W) \\
E'_R(\ell_2,k_1)             & E'_R(\ell_2,k_2)      & \cdots    & E'_R(\ell_2,k_W)\\
    \vdots                      &   \vdots           & \ddots       &   \vdots   \\
E'_R(\ell_U,k_1)             & E'_R(\ell_U,k_2)      & \cdots    & E'_R(\ell_U,k_W)\\
\end{bmatrix}.
\end{equation}
For easier analysis, we assume the adopted frequencies and OAM modes satisfy $k_{w+1}-k_w=1$ and $\ell_{{u}+1}-\ell_{{u}}=1$.

\emph{The aim of 3-D position estimation is to obtain the distance $r^{q_{i}}_t$, the azimuthal angle $\varphi^{q_{i}}_t$ and the elevation angle $\theta^{q_{i}}_t$ of each scattering point of the targets from the elements of the matrix \eqref{matrixE}}.  From \eqref{ER2}, we observe that the azimuthal angle $\varphi^{q_{i}}_t$ is coupled with  the OAM mode number $\ell_u$ and  so is $r^{q_{i}}_t$ with $k_w$. The elevation angle $\theta^{q_{i}}_t$ is associated with both $\ell_u$ and $k_w$. Therefore, we propose to estimate $\{(\theta^{q_{i}}_t,\varphi^{q_{i}}_t)|i=1,2,\cdots,P,q=1,2,\cdots,Q\}$ by  processing the columns of the  matrix \eqref{matrixE},
whose elements depend on $\ell_{u}$, and to estimate $\{(r^{q_{i}}_t,\theta^{q_{i}}_t)|i=1,2,\cdots,P,q=1,2,\cdots,Q\}$ by  processing the row of the matrix \eqref{matrixE}, whose elements depend on $k_{w}$.
To ensure the accuracy of OAM-based target imaging, we assume that  $U\geq QP$ and $W\geq QP$.
Eventually, the estimates of $(\theta^{q_{i}}_{t},\varphi^{q_{i}}_t)$ in the OAM mode domain can be combined with the estimates of $(r^{q_{i}}_t,\theta^{q_{i}}_{t})$ in the frequency domain to estimate the 3-D  position of all the scattering points.
%Therefore, we propose to estimate $\{(\theta^{q_{i}}_t,\varphi^{q_{i}}_t)|i=1,2,\cdots,P,q=1,2,\cdots,Q\}$ by using the internal relationship between $\ell_u$ and $(\theta^{q_{i}}_t,\varphi^{q_{i}}_t)$, and to estimate $\{(r^{q_{i}}_t,\theta^{q_{i}}_t)|i=1,2,\cdots,P,q=1,2,\cdots,Q\}$ by using the internal relationship between $k_w$ and $(r^{q_{i}}_t,\theta^{q_{i}}_t)$. After that, the 3-D estimated position of all the scattering points $\{(\hat{r}^{q_{i}}_t,\hat{\theta}^{q_{i}}_t,\hat{\varphi}^{q_{i}}_t)|i=1,2,\cdots,P,q=1,2,\cdots,Q\}$ can be determined by matching the repeatedly estimated elevation angle $\{\hat{\theta}^{q_{i}}_t|i=1,2,\cdots,P,q=1,2,\cdots,Q\}$.

\vspace{-0.0cm}
\subsubsection{Estimation in OAM Mode Domain}%Estimation of $\theta^{q_{i}}_t$ and $\varphi^{q_{i}}_t$

The MUSIC algorithm, a subspace-based super-resolution algorithm, provides an elegant means for estimating the parameters of complex sinusoidal signals embedded in white Gaussian noise \cite{Schmidt1896Multiple}. In the estimation of $\theta^{q_{i}}_t$ and $\varphi^{q_{i}}_t$, we first denote the $w$-th column of $\mathbf{E}'_R$ as a column vector $\mathbf{e}_w$, i.e.,
\begin{align}
\mathbf{e}_w\!=\!\mathbf{E}'_R(:,w)\!=\![E'_R(\ell_1,k_w),\! E'_R(\ell_2,k_w),\!\cdots, \!E'_R(\ell_U,k_w)]^\mathrm{T},
\end{align}
so that $\mathbf{e}_w$ can be expressed in compact form as
\begin{equation}
\mathbf{e}_w=\mathbf{A}_w\mathbf{s}_w+\mathbf{n}_w,
\end{equation}
where
$\mathbf{A}_w$ $=$ $[e^{i\ell_u\varphi^{q_{i}}_t}{J_{\ell_u}}(k_wR\sin\theta^{q_{i}}_t)]_{U\times QP}$ is the direction matrix containing angle information of all the scattering points, $\mathbf{s}_w$ $=$ $\mathbf{\Sigma}_w \boldsymbol{\sigma}_s$, $\mathbf{\Sigma}_w$ $=$ $\textrm{diag}\{{J_0}(k_wR\sin\theta^{1_{1}}_t)\frac{e^{i2k_wr^{1_{1}}_t}}{({r^{1_{1}}_t})^2},$ $ \cdots,$ ${J_0}(k_wR\sin\theta^{Q_{P}}_t)\frac{e^{i2k_wr^{Q_{P}}_t}}{({r^{Q_{P}}_t})^2}\}$, $\boldsymbol{\sigma}_s$ $=$ $[\sigma_{1_{1}},\cdots,\sigma_{Q_{P}}]^\mathrm{T}$, and $\mathbf{n}_w$ is the  noise vector. In practice, the OAM echo signals of scattering points are usually incoherent, i.e.,
\begin{align}
\begin{split}
\mathbb{E}\left\{\sigma^{}_{q_{i}}\sigma^{*}_{q'_{j}}\right\}= \left \{
\begin{array}{ll}
    \mathbb{E}\left\{\sigma_{q_{i}}\sigma^{*}_{q'_{j}}\right\},                    & {q_{i}} = {q'_{j}},\\
    0,                             & {q_{i}}\neq {q'_{j}}.\\
\end{array}
\right.
\end{split}
\end{align}

Then, the covariance matrix of $\mathbf{e}_w$ can be written as
\begin{align}\label{Rew}
\mathbf{R}_{\mathbf{e}_{w}}=\mathbb{E}\left\{\mathbf{e}_w\mathbf{e}^{\mathrm{H}}_w\right\}=\mathbf{A}_w\mathbf{R}_{\mathbf{s}_{w}}\mathbf{A}_w^{\mathrm{H}}+\mathbf{R}_{\mathbf{n}_w},
\end{align}
where $\mathbf{R}_{\mathbf{s}_{w}}=\mathbf{\Sigma}_w\mathbb{E}\left\{\boldsymbol{\sigma}_s\boldsymbol{\sigma}^\mathrm{H}_s\right\}\mathbf{\Sigma}_w^\mathrm{H}$, $\mathbf{R}_{\mathbf{n}_w}=\mathbb{E}\left\{\mathbf{n}_w\mathbf{n}^{\mathrm{H}}_w\right\}$. The eigen value decomposition (EVD) of $\mathbf{R}_{\mathbf{e}_{w}}$ is
\begin{equation}\label{EVD}
\mathbf{R}_{\mathbf{e}_{w}}=\sum_{u=1}^{U}\lambda^w_u\mathbf{q}^w_u{\mathbf{q}^w_u}^{\mathrm{H}}=\mathbf{Q}_w\mathbf{\Lambda}_w\mathbf{Q}^{\mathrm{H}}_w,
\end{equation}
where $\{\lambda^w_u|u=1,2,\cdots,U\}$ are the eigenvalues of $\mathbf{R}_{\mathbf{e}_{w}}$, $\mathbf{\Lambda}_w$$=$$\textrm{diag}\{\lambda^w_1,$ $\lambda^w_2,\cdots,\lambda^w_U\}$, $\mathbf{q}^w_u$ is the eigenvector corresponding to the eigenvalue $\lambda^w_u$ and $\mathbf{Q}_w= [\mathbf{q}^w_1,\cdots,\mathbf{q}^w_U]$.

In practice, the echo signals and noise are independent of each other, thus, the covariance matrix $\mathbf{R}_{\mathbf{e}_{w}}$ can be decomposed into two mutually orthogonal parts:
\begin{equation}\label{EVD2}
\mathbf{R}_{\mathbf{e}_{w}}=\mathbf{Q}^w_s\mathbf{\Lambda}^w_s{\mathbf{Q}^w_s}^{\mathrm{H}}+\mathbf{Q}^w_n\mathbf{\Lambda}^w_n{\mathbf{Q}^w_n}^{\mathrm{H}},
\end{equation}
where $\mathbf{\Lambda}^w_s$ is the $QP$-dimensional diagonal matrix containing the larger $QP$ eigenvalues of $\mathbf{R}_{\mathbf{e}_{w}}$, $\mathbf{Q}^w_s$ is the $U\times QP$ signal subspace composed of the eigenvectors corresponding to the larger $QP$ eigenvalues, $\mathbf{\Lambda}^w_n$ is the $(U$$-$$QP)$-dimensional diagonal matrix containing the smaller $U$$-$$QP$ eigenvalues of $\mathbf{R}_{\mathbf{e}_{w}}$, $\mathbf{Q}^w_n$ is the $U\times (U$$-$$QP)$ noise subspace composed of the eigenvectors corresponding to the smaller $U$$-$$QP$ eigenvalues.

Based on \eqref{Rew} and \eqref{EVD2}, we can see that
\begin{align}\label{EVD3}
\mathbf{R}_{\mathbf{e}_{w}}\mathbf{Q}^w_n=\mathbf{A}_w\mathbf{R}_{\mathbf{s}_{w}}\mathbf{A}_w^{\mathrm{H}}\mathbf{Q}^w_n+\mathbf{R}_{\mathbf{n}_w}\mathbf{Q}^w_n=\mathbf{R}_{\mathbf{n}_w}\mathbf{Q}^w_n,
\end{align}
i.e., $\mathbf{A}_w\mathbf{R}_{\mathbf{s}_{w}}\mathbf{A}_w^{\mathrm{H}}\mathbf{Q}^w_n$ $=$ $0$. Since $\mathbf{R}_{\mathbf{s}_{w}}$ $=$ $\mathbf{\Sigma}_w\mathbb{E}\left\{\boldsymbol{\sigma}_s\boldsymbol{\sigma}^\mathrm{H}_s\right\}\mathbf{\Sigma}_w^\mathrm{H}$ is a full rank matrix, which must be reversible, $\mathbf{A}_w^{\mathrm{H}}\mathbf{Q}^w_n=0$, that is, the direction matrix $\mathbf{A}_w$ is orthogonal to the noise subspace $\mathbf{Q}^w_n$.

Based on the orthogonality between $\mathbf{A}_w$ and $\mathbf{Q}^w_n$, the spatial-spectral function can be constructed as
\begin{equation}\label{Pmusic}
P_w(\theta_t,\varphi_t)=\frac{1}{\mathbf{a}^\mathrm{H}_w(\theta_t,\varphi_t)\mathbf{Q}^w_n{\mathbf{Q}^w_n}^{\mathrm{H}}\mathbf{a}_w(\theta_t,\varphi_t)},
\end{equation}
where $\mathbf{a}_w(\theta_t,\varphi_t)$$=$$[e^{i\ell_1\varphi_t}{J_{\ell_1}}(k_wR\sin\theta_t),$$e^{i\ell_2\varphi_t}{J_{\ell_2}}(k_wR\sin\theta_t),$ $\cdots,$ $e^{i\ell_U\varphi_t}{J_{\ell_U}}(k_wR\sin\theta_t)]^\mathrm{T}$, $\theta_t\in [0, \pi]$, $\varphi_t\in [0, 2\pi]$. Then, the estimates $\{(\hat{\theta}^{q_{i}}_{t,w},\hat{\varphi}^{q_{i}}_{t,w})|i=1,2,\cdots,P,q=1,2,\cdots,Q\}$ can be obtained through the 2-D spectrum peak searching in \eqref{Pmusic}.

Hence, there are $W$ estimates of $\{(\theta^{q_{i}}_t,\varphi^{q_{i}}_t)|i$ $=$ $1,$ $2,\cdots,P,q=1,2,\cdots,Q\}$ in total, which can be expressed as
\begin{align}
\hat{\theta}^{q_{i}}_{t,w}=\theta^{q_{i}}_t+\varepsilon^{{q_{i}}}_{\theta,w},
\hat{\varphi}^{q_{i}}_{t,w}=\varphi^{q_{i}}_t+\varepsilon^{{q_{i}}}_{\varphi,w},
\end{align}
where $\varepsilon^{{q_{i}}}_{\theta,w}$ and $\varepsilon^{{q_{i}}}_{\varphi,w}$ represent the estimation errors, $i$ $=$ $1,$ $2,$ $\cdots,$ $P$, $q$ $=$ $1,$ $2,$ $\cdots,$ $Q$, $w$ $=$ $1,$ $2,$ $\cdots,$ $W$. Suppose that $\{\varepsilon^{{q_{i}}}_{\theta,w}\}$ %|w$ $=$ $1,2,\cdots,W\}$
and $\{\varepsilon^{{q_{i}}}_{\varphi,w}\}$ %|w$ $=$ $1,2,\cdots,W\}$
have the same average variance $\textrm{Var}(\varepsilon^{q_{i}}_{\theta})$ and $\textrm{Var}(\varepsilon^{q_{i}}_{\varphi})$, respectively. Accordingly, since it is
$\textrm{Var}\left(\frac{1}{W}\sum_{w=1}^{W}\hat{\varphi}^{q_{i}}_{t,w}\right)= \frac{\textrm{Var}\left(\varepsilon^{q_{i}}_{\varphi}\right)}{W}$,
%
%Therefore, $\hat{\theta}^{q_{i}}_t=\frac{1}{W}\sum_{w=1}^{W}\hat{\theta}^{q_{i}}_{t,w}$ and
we can adopt
\begin{equation}
\hat{\varphi}^{q_{i}}_t=\frac{1}{W}\sum_{w=1}^{W}\hat{\varphi}^{q_{i}}_{t,w}
\end{equation}
%
% are respectively adopted
as the estimate of
%$\theta^{q_{i}}_t$ and
 $\varphi^{q_{i}}_t$.

\begin{algorithm}[t]
\label{alg1}
\caption{OAM-based 3-D Position Esitmation}
\hspace*{0.02in} {\bf Input:}
$\mathbf{E}'_R$\\
\hspace*{0.02in} {\bf Output:}
$\{(\hat{r}^{q_{i}}_t,\hat{\theta}^{q_{i}}_t,\hat{\varphi}^{q_{i}}_t)|i=1,2,\cdots,P,q=1,2,\cdots,Q\}$
\begin{algorithmic}[1]
\State \textbf{procedure}
\State$\mathbf{e}_w\leftarrow\mathbf{E}'_R(:,w), w=1,2,\cdots,W$
\State$\mathbf{R}_{\mathbf{e}_{w}}\leftarrow\mathbb{E}\left\{\mathbf{e}_w\mathbf{e}^{\mathrm{H}}_w\right\}, w=1,2,\cdots,W$
\For{$w = 1 \to W$}\\
$\mathbf{Q}_w$, $\mathbf{\Lambda}_w\leftarrow$ decompose $\mathbf{R}_{\mathbf{e}_{w}}$ such that $\mathbf{Q}_w\mathbf{\Lambda}_w\mathbf{Q}^{\mathrm{H}}_w$\\
$\mathbf{\Lambda}_w\leftarrow$ $\textrm{diag}\{\lambda^w_{1},$$\cdots,$$\lambda^w_{U}\}$, $\lambda^w_{1}$$\geq$$\cdots$$\geq$$\lambda^w_{ QP}$$\geq$$\cdots$$\geq$$\lambda^w_{U}$\\
$\mathbf{\Lambda}^w_n$ $\leftarrow$ $\textrm{diag}\{\lambda^w_{QP+1},$$\cdots,$$\lambda^w_{U}\}$\\
$\mathbf{Q}^w_n$ $\leftarrow$ the column of $\mathbf{Q}_w$ corresponding to $\mathbf{\Lambda}^w_n$\\
$\mathbf{a}_w(\theta_t,\varphi_t)\leftarrow$ $[e^{i\ell_u\varphi_t}{J_{\ell_u}}(k_wR\sin\theta_t)]_{U\times 1}$\\
$P_w(\theta_t,\varphi_t)$ $\leftarrow$ $1/({\mathbf{a}^\mathrm{H}_w(\theta_t,\varphi_t)\mathbf{Q}^w_n{\mathbf{Q}^w_n}^{\mathrm{H}}\mathbf{a}_w(\theta_t,\varphi_t))}$\\
$(\hat{\theta}^{q_{i}}_{t,w},\hat{\varphi}^{q_{i}}_{t,w})$ $\leftarrow$ 2-D spectrum peak searching
in $P_w(\theta_t,\varphi_t)$\\
\textbf{end for}
\EndFor
%\State$\hat{\theta}^{q_{i}}_t \leftarrow$ $\frac{1}{W}\sum_{w=1}^{W}\hat{\theta}^{q_{i}}_{t,w}$, $i=1,\cdots,P$, $q=1,\cdots,Q$
\State$\hat{\varphi}^{q_{i}}_t \leftarrow$ $\frac{1}{W}\sum_{w=1}^{W}\hat{\varphi}^{q_{i}}_{t,w}$, $i=1,\cdots,P$, $q=1,\cdots,Q$
\State$\mathbf{e}_u\leftarrow\mathbf{E}'_R(u,:), u=1,2,\cdots,U$
\State$\mathbf{R}_{\mathbf{e}_{u}}\leftarrow\mathbb{E}\left\{\mathbf{e}_u\mathbf{e}^{\mathrm{H}}_u\right\}, u=1,2,\cdots,U$
\For{$u = 1 \to U$}\\
$\mathbf{Q}_u$, $\mathbf{\Lambda}_u\leftarrow$ decompose $\mathbf{R}_{\mathbf{e}_{u}}$ such that $\mathbf{Q}_u\mathbf{\Lambda}_u\mathbf{U}^{\mathrm{H}}_u$\\
$\mathbf{\Lambda}_u\leftarrow$ $\textrm{diag}\{\lambda^u_{1},$$\cdots,$$\lambda^u_{W}\}$,$\lambda^u_{1}$$\geq$$\cdots$$\geq$$\lambda^u_{QP}$$\geq$$\cdots$$\geq$$\lambda^u_{W}$\\
$\mathbf{\Lambda}^u_n$ $\leftarrow$ $\textrm{diag}\{\lambda^u_{QP+1},$$\cdots,$$\lambda^u_{W}\}$\\
$\mathbf{Q}^u_n$ $\leftarrow$ the column of $\mathbf{Q}_u$ corresponding to $\mathbf{\Lambda}^u_n$\\
$\mathbf{b}_u(r_t,\theta_t)\leftarrow$ $[e^{i2k_wr_t}{J_{\ell_u}}(k_wR\sin\theta_t){J_{0}}(k_wR\sin\theta_t)]_{W\times 1}$\\
$P_u(r_t,\theta_t)$ $\leftarrow$ $1/({\mathbf{b}^\mathrm{H}_u(r_t,\theta_t)\mathbf{Q}^u_n{\mathbf{Q}^u_n}^{\mathrm{H}}\mathbf{b}_u(r_t,\theta_t))}$\\
$(\hat{r}^{q_{i}}_{t,u},\hat{\theta}^{q_{i}}_{t,u})$ $\leftarrow$ 2-D spectrum peak searching
in $P_u(r_t,\theta_t)$\\
\textbf{end for}
\EndFor
\State$\hat{r}^{q_{i}}_t \leftarrow$ $\frac{1}{U}\sum_{u=1}^{U}\hat{r}^{q_{i}}_{t,u}$, $i=1,\cdots,P$, $q=1,\cdots,Q$
%\State$\hat{\theta}'^{q_{i}}_t \leftarrow$ $\frac{1}{U}\sum_{u=1}^{U}\hat{\theta}^{q_{i}}_{t,u}$, $i=1,\cdots,P$, $q=1,\cdots,Q$
%\State$(\hat{r}^{q_{i}}_t,\hat{\theta}^{q_{i}}_t,\hat{\varphi}^{q_{i}}_t) \leftarrow$ \textcolor{blue}{pairing $(\hat{\theta}^{q_{i}}_t,\hat{\varphi}^{q_{i}}_t)$ with $(\hat{r}^{q_{i}}_t,\hat{\theta}'^{q_{i}}_t)$ by comparing $\hat{\theta}^{q_{i}}_t$ and $\hat{\theta}'^{q_{i}}_t$}
\State$\hat{\theta}^{q_{i}}_t \leftarrow$ $\frac{1}{U+W}\left(\sum_{w=1}^{W}\hat{\theta}^{q_{i}}_{t,w}+\sum_{u=1}^{U}\hat{\theta}^{q_{i}}_{t,u}\right)$, $i=1,\cdots,P$, $q=1,\cdots,Q$
\State \textbf{end procedure}
\end{algorithmic}
\end{algorithm}
%matching by $\hat{\theta}^{q_{i}}_t$ and $\hat{\theta}'^{q_{i}}_t$
%

\vspace{0.0cm}
\subsubsection{Estimation in Frequency Domain}%Estimation of $r^{q_{i}}_t$ and $\theta^{q_{i}}_t$

In the estimation of $r^{q_{i}}_t$ and $\theta^{q_{i}}_t$, we similarly denote the $u$-th row of $\mathbf{E}'_R$ as a column vector $\mathbf{e}_u$, i.e.,
\begin{align}
\mathbf{e}_u\!=\!\mathbf{E}'_R(u,:)\!=\![E'_R(\ell_u,k_1),\! E'_R(\ell_u,k_2),\!\cdots,\! E'_R(\ell_u,k_W)]^\mathrm{T},
\end{align}
so that $\mathbf{e}_u$ can be expressed in compact form as
\begin{equation}
\mathbf{e}_u=\mathbf{B}_u\mathbf{s}_u+\mathbf{n}_u,
\end{equation}
where
$\mathbf{B}_u$ $=$ $[e^{i2k_wr^{q_{i}}_t}{J_{\ell_u}}(k_wR\sin\theta^{q_{i}}_t){J_{0}}(k_wR\sin\theta^{q_{i}}_t)]_{W\times QP}$ is the position matrix containing 2-D position information of  all the scattering points, $\mathbf{s}_u$ $=$ $\mathbf{\Sigma}_u \boldsymbol{\sigma}_s$, $\mathbf{\Sigma}_u$ $=$ $\textrm{diag}\{\frac{e^{i\ell_u\varphi^{1_{1}}_t}}{({r^{1_{1}}_t})^2},$ $\cdots,$$\frac{e^{i\ell_u\varphi^{Q_{P}}_t}}{({r^{Q_{P}}_t})^2}\}$, and $\mathbf{n}_u$ is the noise vector. Then, by computing the correlation matrix $\mathbf{R}_{\mathbf{e}_{u}}=\mathbb{E}\left\{\mathbf{e}_u\mathbf{e}^{\mathrm{H}}_u\right\}$ and,
following the same method outlined in the previous section,
% to estimate $\{(\theta^{q_{i}}_t,\varphi^{q_{i}}_t)|$ $i=1,2,\cdots,P,q=1,2,\cdots,Q\}$,
we can obtain the estimates $\{(\hat{r}^{q_{i}}_{t,u},\hat{\theta}^{q_{i}}_{t,u})|$$i$ $=$ $1,$ $2,$ $\cdots,$ $P,$ $q$ $=$ $1,$ $2,$ $\cdots,$ $Q,$ $u$ $=$ $1,$ $2,$ $\cdots,$ $U\}$,
so that eventually it is
\begin{equation}
\hat{r}^{q_{i}}_t=\frac{1}{U}\sum_{u=1}^{U}\hat{r}^{q_{i}}_{t,u}.
\end{equation}
and
\begin{equation}
\hat{\theta}^{q_{i}}_t=\frac{1}{U+W}\left(\sum_{w=1}^{W}\hat{\theta}^{q_{i}}_{t,w}+\sum_{u=1}^{U}\hat{\theta}^{q_{i}}_{t,u}\right).
\end{equation}
The detailed procedure to  obtain the 3-D estimated position $(\hat{r}^{q_{i}}_t,\hat{\theta}^{q_{i}}_t,\hat{\varphi}^{q_{i}}_t)$ of the $q_i$ scattering points is summarized in Algorithm 1.
 %\textcolor{blue}{After that, by comparing the repeatedly estimated elevation angles $\hat{\theta}^{q_{i}}_t$ and $\hat{\theta}'^{q_{i}}_t$, the estimates $(\hat{\theta}^{q_{i}}_t,\hat{\varphi}^{q_{i}}_t)$ can be combined with the estimates $(\hat{r}^{q_{i}}_t,\hat{\theta}'^{q_{i}}_t)$ to finally obtain the 3-D estimated position $(\hat{r}^{q_{i}}_t,\hat{\theta}^{q_{i}}_t,\hat{\varphi}^{q_{i}}_t)$ of the $q_i$ scattering points}, and the detailed procedure is summarized in Algorithm 1.

\vspace{-0.2cm}
\subsection{OAM-based Rotation Velocity Detection}

\begin{figure}[t] %figure1
\setlength{\abovecaptionskip}{-0.2cm}   %è°æŽåŸçæ é¢äžåŸè·çŠ»
\setlength{\belowcaptionskip}{-0.4cm}   %è°æŽåŸçæ é¢äžäžæè·çŠ»
\footnotesize
\begin{center}
\includegraphics[width=6.4cm,height=5.7cm]{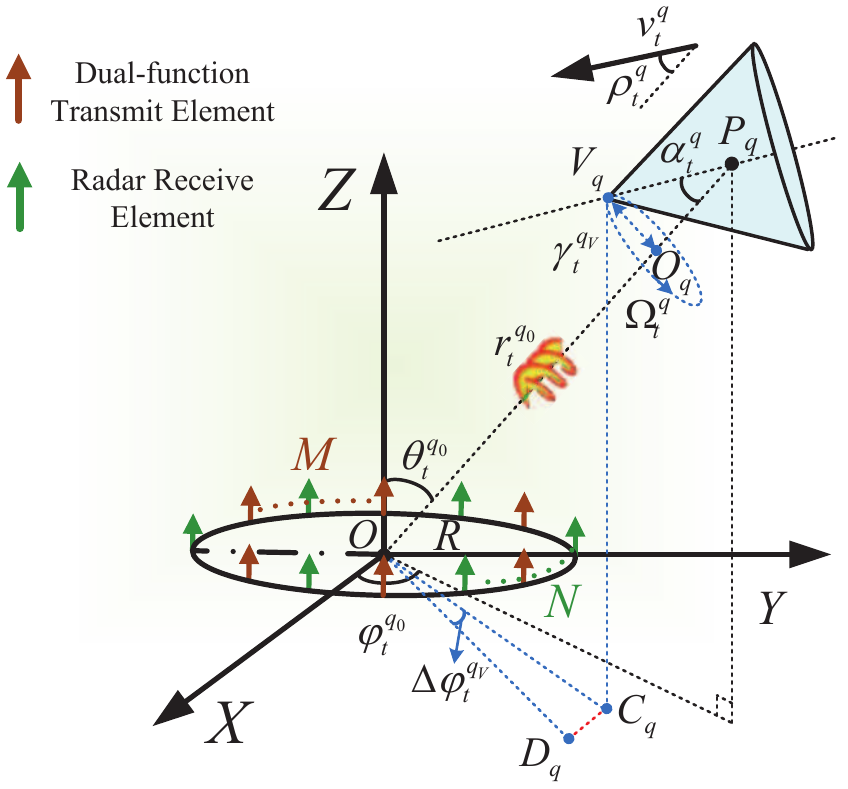}
\end{center}
\caption{The diagram of the OAM-based rotation velocity detection.}
\label{Fig3}
\end{figure}

\vspace{-0.0cm}
\subsubsection{Problem Formulation}

\vspace{-0.0cm}
The precession characteristic of targets is the key basis for distinguishing the type of targets \cite{Gao2010Doppler,Lei2012Micromotion}. In the scenario shown in Fig.\ref{Fig1}, the targets are moving along the trajectory while rotating around their axis, and the rotation velocity is an important precession parameter. In this paper, we will focus on the OAM-based rotation velocity detection to lay a foundation for distinguishing the type of targets.

For easier understanding, we take the vertex $\text{V}_q$ of $q$-th target as an example to derive the theoretical expression of the rotation velocity. The motion state of the $q$-th target is shown in Fig.\ref{Fig3}, whose centroid at time instant $t$ has coordinates $(r^{q_{0}}_t,\theta^{q_{0}}_t,\varphi^{q_{0}}_t)$. The $q$-th target is moving in the direction $\rho^q_t$ at a velocity $v^q_t$  while rotating around the $\text{O}\text{P}_q$-axis at an angular velocity $\Omega^q_t$. Suppose the rotation center of $\text{V}_q$  is $\text{O}_q$ and the rotation radius is $\gamma_t^{q_{_V}}$, the vector $\overrightarrow{\text{O}_q\text{V}_q}$ can be written as
\begin{equation}\label{P1P2}
\overrightarrow{\text{O}_q\text{V}_q}=\begin{pmatrix}
x^V_t\\
y^V_t\\
z^V_t\\
\end{pmatrix}
=\mathbf{R}_{\text{ro}}\cdot
\begin{small}
\begin{pmatrix}
\gamma_t^{q_{_V}}\cos(\Omega^{q}_tt+\psi^{q_{_V}}_0)\\
\gamma_t^{q_{_V}}\sin(\Omega^{q}_tt+\psi^{q_{_V}}_0)\\
0\\
\end{pmatrix}\end{small},
\end{equation}
where
\begin{equation*}
\mathbf{R}_{\text{ro}}=
\begin{small}\begin{bmatrix}
\cos\varphi^{q_{0}}_t                         & -\sin\varphi^{q_{0}}_t                   & 0\\
\cos\theta^{q_{0}}_t\sin\varphi^{q_{0}}_t           & \cos\theta^{q_{0}}_t\cos\varphi^{q_{0}}_t      & -\sin\theta^{q_{0}}_t\\
\sin\theta^{q_{0}}_t\sin\varphi^{q_{0}}_t           & \sin\theta^{q_{0}}_t\cos\varphi^{q_{0}}_t      & \cos\theta^{q_{0}}_t\\
\end{bmatrix}\end{small}
\end{equation*}
is the rotation matrix determined by the the position of the $q$-th target and $\psi^{q_{_V}}_0$ is the initial azimuth angle of the vertex $\text{V}_q$. When $\text{V}_q$ rotates around $\text{O}\text{P}_q$-axis, the corresponding linear velocity vector $\boldsymbol{v}_{_V}$ is perpendicular to $\overrightarrow{\text{O}\text{P}_q}$ and $\overrightarrow{\text{O}_q\text{V}_q}$, thus, the linear velocity vector $\boldsymbol{v}_{_V}$ can be expressed as
%$\psi^q_0$ is the initial azimuth angle of the vertex $\text{V}_q$ in the coordinate system $\text{Z}-\text{XOY}$
\begin{align}
\boldsymbol{v}_{_V}&=(v^x_t,v^y_t,v^z_t)=\hat{\mathbf{r}}^{q_{0}}_t\times\overrightarrow{\text{O}_q\text{V}_q}\nonumber\\
&=(y^{q_{0}}_tz^V_t\!\!-\!z^{q_{0}}_ty^V_t,z^{q_{0}}_tx^V_t\!\!-\!x^{q_{0}}_tz^V_t,x^{q_{0}}_ty^V_t\!\!-\!y^{q_{0}}_tx^V_t),
\end{align}
where $\hat{\mathbf{r}}^{q_{0}}_t$ $=$ $\mathbf{r}^{q_{0}}_t/|\mathbf{r}^{q_{0}}_t|$ $=$ $(x^{q_{0}}_t,$ $y^{q_{0}}_t,$ $z^{q_{0}}_t)^{\mathrm{T}}$ $=$ $(\sin\theta^{q_{0}}_t\cos\varphi^{q_{0}}_t,$ $\sin\theta^{q_{0}}_t\sin\varphi^{q_{0}}_t,$ $\cos\theta^{q_{0}}_t)^{\mathrm{T}}$ is the unit vector of $\overrightarrow{\text{O}\text{P}_q}$. Then, within the period $\Delta t$, the moving distance of $C_q$ that is the projection of the vertex $\text{V}_q$ in the $\text{XOY}$-plane can be expressed as
\begin{align}\label{dp}
&\Delta d^{q_{_V}}_t=(\boldsymbol{v}_{_V}\cdot\hat{\mathbf{r}}^{q_{0}}_{\bot})\Delta t=\frac{-y^{q_{0}}_tv^x_t+x^{q_{0}}_tv^y_t}{\sqrt{(x^{q_{0}}_t)^2+(y^{q_{0}}_t)^2}}\Delta t\nonumber\\
&=\!\!-\gamma_t^{q_{_V}}\!\Omega^q_t\!\sqrt{\!\sin^2\!\varphi^{q_{0}}_t\!\!+\!\cos^2\!\varphi^{q_{0}}_t\!\sin^2\!\theta^{q_{0}}_t}\cos(\Omega^q_t t\!+\!\psi^{q_{_V}}_0\!\!+\!\delta^{q_{0}}_t)\Delta t,
\end{align}
where $\hat{\mathbf{r}}^{q_{0}}_{\bot}$ $=$ $\frac{1}{\sqrt{(x^{q_{0}}_t)^2+(y^{q_{0}}_t)^2}}(-y^{q_{0}}_t,x^{q_{0}}_t,0)^{\mathrm{T}}$ is the unit vector on the $\text{XOY}$-plane projected by the vector perpendicular to $\overrightarrow{\text{O}\text{P}_q}$, $\delta^{q_{0}}_t=\arctan(\cos\varphi^{q_{0}}_t\sin\theta^{q_{0}}_t/\sin\varphi^{q_{0}}_t)$. After that, the change of the azimuth angle of the vertex $\text{V}_q$ can  be expressed as
\begin{align}\label{phip}
&\Delta\varphi^{q_{_V}}_t=\lim_{\Delta t \to 0}\frac{\Delta d^{q_{_V}}_t}{|\overrightarrow{\text{O}\text{C}_q}|}.
\end{align}
With \eqref{phip}, we can derive the theoretical rotational Doppler shift of the vertex $\text{V}_q$ induced by azimuthal change.

\begin{figure}[t] %figure1
\setlength{\abovecaptionskip}{-0.2cm}   %è°æŽåŸçæ é¢äžåŸè·çŠ»
\setlength{\belowcaptionskip}{-0.5cm}   %è°æŽåŸçæ é¢äžäžæè·çŠ»
\footnotesize
\begin{center}
\includegraphics[width=7.8cm,height=10.4cm]{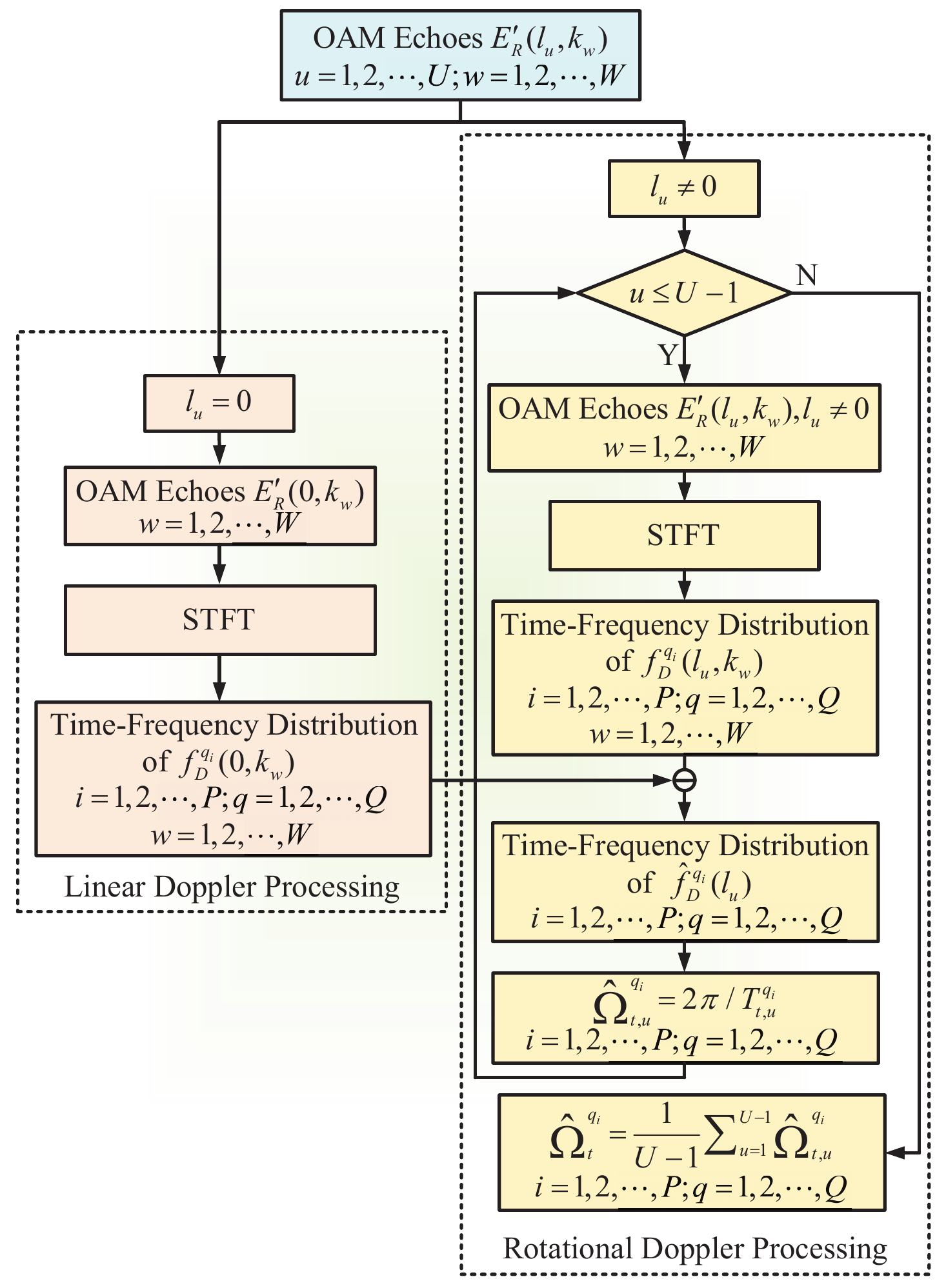}
\end{center}
\caption{The flowchart of the OAM-based rotation velocity detection.}
\label{Fig4}
\end{figure}

Based on \eqref{ER2}, the phase term of the OAM echo signal of the vertex $\text{V}_q$ on the $u$-th mode at the $w$-th subcarrier can be written as
\begin{align}
\Phi^{q_{_V}}_t(\ell_u,k_w)=2k_wr^{q_{_V}}_t+\ell_u\varphi^{q_{_V}}_t.
\end{align}
Thus, the Doppler shift of the vertex $\text{V}_q$ can be expressed as
\begin{align}\label{Doppler}
f^{q_{_V}}_D(\ell_u,k_w)\!=\!\!\frac{1}{2\pi}\!\cdot\!\frac{\text{d}\Phi^{q_{_V}}_t(\ell_u,k_w)}{\text{d}t}\!=\!\!f^{q_{_V}}_L(k_w)\!+\!f^{q_{_V}}_{\Omega}(\ell_u),
\end{align}
where
\begin{align}\label{LDoppler}
f^{q_{_V}}_L(k_w)=\frac{1}{2\pi}k_wv^q_t\cos\rho^q_t
\end{align}
is the linear Doppler shift induced by distance variation, and
\begin{align}\label{RDoppler1}
f^{q_{_V}}_{\Omega}(\ell_u)\!=\!\frac{\ell_u}{2\pi}\cdot\lim_{\Delta t \to 0}\frac{\Delta\varphi^{q_{_V}}_t}{\Delta t}\!=\!\frac{\ell_u}{2\pi}g^q_t\Omega^q_t\cos(\Omega^q_t t+\psi^{q_{_V}}_0+\delta^{q_{0}}_t)
\end{align}
is the rotational Doppler shift induced by azimuthal change, and $g^q_t$ $=$ $\frac{-\gamma_t^{q_{_V}}\sqrt{\sin^2\varphi^{q_{0}}_t+\cos^2\varphi^{q_{0}}_t\sin^2\theta^{q_{0}}_t}}{(r^{q_{0}}_t-\gamma_t^{q_{_V}}/\tan\alpha^q_t)\sin\theta^{q_{0}}_t}$.
%$g^q_t$ $=$ $\frac{-\gamma_t^q\Omega^q_t\sqrt{\sin^2\varphi^q_t+\cos^2\varphi^q_t\sin^2\theta^q_t}}{(r^q_t-\gamma_t^q/\tan\alpha^q_t)\sin\vartheta^q_t}$, $\vartheta^q_t$ is the initial elevation angle of the vertex $\text{V}_q$ in the coordinate system $\text{Z}-\text{XOY}$.

From \eqref{RDoppler1}, we can observe that the rotational Doppler shift of $\text{V}_q$ is frequency independent and determined only by the OAM mode number $\ell_u$ and angular velocity $\Omega^q_t$, so that  the higher the OAM mode is, the larger is the rotational Doppler shift. In the next part, we will discuss in detail how to extract the rotational Doppler shift of each scattering point from \eqref{ER2} and obtain the angular velocity $\Omega^q_t$.

\vspace{0.0cm}
\subsubsection{Detection of $\Omega^q_t$}

The rotation of a rigid body is essentially a kind of non-uniform motion. Thus, the echo signals are non-linear and non-stationary \cite{Wang2021Detection}. The core problem of extracting and analyzing the rotational Doppler shift of each scattering point from \eqref{ER2} is the processing of time-varying signals. Due to its suitability for non-stationary signal analysis, the short-time Fourier transform (STFT) method has long been used for Doppler imaging \cite{Chen1998Joint}. After time-frequency processing of the OAM echo signal $E'_R(\ell_u,k_w)$, we can obtain the time-frequency distribution of each scattering point superposed by the linear Doppler shift and the rotational Doppler shift. Then, the time-frequency distribution of the $q_i$-th scattering point only consisting of the rotational Doppler shift can be obtained as follows:
%by offsetting the linear Doppler shift using $f^{q_{i}}_D(0,k_w)$, i.e.,
%the time-frequency distribution of
\begin{align}\label{RDoppler2}
\setcounter{equation}{29}
\hat{f}^{q_{i}}_{D}(\ell_u)&=f^{q_{i}}_D(\ell_u,k_w)-f^{q_{i}}_D(0,k_w), \ \ell_u\neq0.
\end{align}
%
%\begin{align}\label{RDoppler2}
%\setcounter{equation}{27}
%\hat{f}^{q_{i}}_{D}(\ell_u)&=f^{q_{i}}_D(\ell_u,k_w)-f^{q_{i}}_D(0,k_w)\nonumber\\
%&=\frac{\ell_u}{2\pi}g^q_t\Omega^q_t\cos(\Omega^q_t t+\psi^q_0+\delta^q_t), \ \ell_u\neq0.
%\end{align}
%
After that, the rotation period $T^{q_{i}}_{t,u}$ of $q_i$-th scattering point can be obtained from the time-frequency distribution of $\hat{f}^{q_{i}}_{D}(\ell_u)$, and the angular velocity can be calculated as $\hat{\Omega}^{q_{i}}_{t,u}=2\pi/T^{q_{i}}_{t,u}$.

Hence, there are $U-1$ estimates of $\{\Omega^{q_i}_t|$$i$$=$$1,$ $2,$ $\cdots,$ $P,$ $q$ $=$ $1,$ $2,$ $\cdots,$ $Q\}$ in total, which can be expressed as
\begin{align}
\hat{\Omega}^{q_{i}}_{t,u}=\Omega^{q_{i}}_t+\varepsilon^{q_{i}}_{\Omega,u},
\end{align}
where $\varepsilon^{q_{i}}_{\Omega,u}$ represents the estimation errors, $i$ $=$ $1,$ $2,$ $\cdots,$ $P$, $q$ $=$ $1,$ $2,$ $\cdots,$ $Q$, $u$ $=$ $1,$ $2,$ $\cdots,$ $U-1$. Suppose that $\{\varepsilon^{q_{i}}_{\Omega,u}|$ $u$ $=$ $1,2,\cdots,U-1\}$ have the same average variance $\textrm{Var}(\varepsilon^{q_{i}}_{\Omega})$, thus,
\begin{align}
\textrm{Var}\left(\frac{1}{U-1}\sum_{u=1}^{U-1}\hat{\Omega}^{q_{i}}_{t,u}\right)= \frac{\textrm{Var}\left(\varepsilon^{q_{i}}_{\Omega}\right)}{U-1}.
\end{align}
Therefore, $\hat{\Omega}^{q_{i}}_t=\frac{1}{U-1}\sum_{u=1}^{U-1}\hat{\Omega}^{q_{i}}_{t,u}$ is adopted as the estimate of $\Omega^{q_{i}}_t$, and the detailed procedure is summarized in Fig.\ref{Fig4}. It should be noted that the detected rotation velocity of the scattering points in the same target should be the same, except that at the centroid of the target, its rotation velocity is detected as zero since its rotation Doppler shift is zero.

\begin{figure*}\label{Fisher}
\setcounter{equation}{34}
\setlength{\abovecaptionskip}{0.0cm}   %è°æŽåŸçæ é¢äžåŸè·çŠ»
\setlength{\belowcaptionskip}{-0.4cm}   %è°æŽåŸçæ é¢äžäžæè·çŠ»
\begin{align*}\label{Fisher}
\mathbf{J}&\!=\!\mathbb{E}\!\left\{\!\left[\frac{\partial}{\partial\boldsymbol{\vartheta}}\ln f(\mathbf{E}_R,\boldsymbol{\vartheta})\right]\!\!\!\left[\frac{\partial}{\partial\boldsymbol{\vartheta}}\ln f(\mathbf{E}_R,\boldsymbol{\vartheta})\right]^\mathrm{T}\!\right\}
\!=\!\mathbb{E}\!\left\{\!\left[\frac{\partial}{\partial\boldsymbol{\vartheta}}\ln f(\mathbf{E}_R|\boldsymbol{\vartheta})\right]\!\!\!\left[\frac{\partial}{\partial\boldsymbol{\vartheta}}\ln f(\mathbf{E}_R|\boldsymbol{\vartheta})\right]^\mathrm{T}\!\right\}
\!=\!\!-\mathbb{E}\!\left\{\!\frac{\partial}{\partial\boldsymbol{\vartheta}}\!\!\left[\frac{\partial}{\partial\boldsymbol{\vartheta}}\ln f(\mathbf{E}_R|\boldsymbol{\vartheta})\right]^\mathrm{T}\!\right\}\nonumber\\
&=\frac{1}{\xi^2}
\begin{bmatrix}
\sum\limits_{u=1}^{U}\!\sum\limits_{w=1}^{W}\!\!\textrm{Re}\!\bigg[\!\frac{\partial p(\ell_u,k_w,\boldsymbol{\vartheta})}{\partial r^{1_0}_t}\!\frac{\partial p(\ell_u,k_w,\boldsymbol{\vartheta})}{\partial r^{1_0}_t}\!\bigg]             & \sum\limits_{u=1}^{U}\!\sum\limits_{w=1}^{W}\!\!\textrm{Re}\!\bigg[\!\frac{\partial p(\ell_u,k_w,\boldsymbol{\vartheta})}{\partial r^{{1_0}}_t}\!\frac{\partial p(\ell_u,k_w,\boldsymbol{\vartheta})}{\partial \theta^{{1_0}}_t}\!\bigg]        & \cdots & \sum\limits_{u=1}^{U}\!\sum\limits_{w=1}^{W}\!\!\textrm{Re}\!\bigg[\!\frac{\partial p(\ell_u,k_w,\boldsymbol{\vartheta})}{\partial r^{{1_0}}_t}\!\frac{\partial p(\ell_u,k_w,\boldsymbol{\vartheta})}{\partial \Omega^{{Q_{_V}}}_t}\!\bigg] \\
\sum\limits_{u=1}^{U}\!\sum\limits_{w=1}^{W}\!\!\textrm{Re}\!\bigg[\!\frac{\partial p(\ell_u,k_w,\boldsymbol{\vartheta})}{\partial \theta^{1_{0}}_t}\!\frac{\partial p(\ell_u,k_w,\boldsymbol{\vartheta})}{\partial r^{1_0}_t}\!\bigg]             & \sum\limits_{u=1}^{U}\!\sum\limits_{w=1}^{W}\!\!\textrm{Re}\!\bigg[\!\frac{\partial p(\ell_u,k_w,\boldsymbol{\vartheta})}{\partial \theta^{1_0}_t}\!\frac{\partial p(\ell_u,k_w,\boldsymbol{\vartheta})}{\partial \theta^{1_0}_t}\!\bigg]     & \cdots
& \sum\limits_{u=1}^{U}\!\sum\limits_{w=1}^{W}\!\!\textrm{Re}\!\bigg[\!\frac{\partial p(\ell_u,k_w,\boldsymbol{\vartheta})}{\partial \theta^{1_0}_t}\!\frac{\partial p(\ell_u,k_w,\boldsymbol{\vartheta})}{\partial \Omega^{{Q_{_V}}}_t}\!\bigg]\\
    \vdots                      &   \vdots           & \ddots       &   \vdots   \\
\sum\limits_{u=1}^{U}\!\sum\limits_{w=1}^{W}\!\!\textrm{Re}\!\bigg[\!\frac{\partial p(\ell_u,k_w,\boldsymbol{\vartheta})}{\partial \Omega^{Q_{_V}}_t}\!\frac{\partial p(\ell_u,k_w,\boldsymbol{\vartheta})}{\partial r^{1_0}_t}\!\bigg]            & \sum\limits_{u=1}^{U}\!\sum\limits_{w=1}^{W}\!\!\textrm{Re}\!\bigg[\!\frac{\partial p(\ell_u,k_w,\boldsymbol{\vartheta})}{\partial \Omega^{Q_{_V}}_t}\!\frac{\partial p(\ell_u,k_w,\boldsymbol{\vartheta})}{\partial \theta^{1_0}_t}\!\bigg]     & \cdots    & \sum\limits_{u=1}^{U}\!\sum\limits_{w=1}^{W}\!\!\textrm{Re}\!\bigg[\!\frac{\partial p(\ell_u,k_w,\boldsymbol{\vartheta})}{\partial \Omega^{Q_{_V}}_t}\!\frac{\partial p(\ell_u,k_w,\boldsymbol{\vartheta})}{\partial \Omega^{Q_{_V}}_t}\!\bigg]\\
\end{bmatrix},
\end{align*}
where\\
\begin{align*}
\frac{\partial p(\ell_u,k_w,\boldsymbol{\vartheta})}{\partial r^{q_{0}}_t}=&-j\frac{A_u\mu_0\omega_w d^2MN}{4\pi}i^{-\ell_u}\sigma_{q_{0}}\frac{2e^{i2k_wr^{q_{0}}_t}}{{(r^{q_{0}}_t)}^2}e^{i\ell_u\varphi^{q_{0}}_t}{J_{\ell_u}}(k_wR\sin\theta^{q_{0}}_t){J_0}(k_wR\sin\theta^{q_{0}}_t)(ik_w-\frac{1}{r^{q_{0}}_t})s(\ell_u,k_w),
\end{align*}
\begin{align*}
\frac{\partial p(\ell_u,k_w,\boldsymbol{\vartheta})}{\partial \theta^{q_{0}}_t}=&-j\frac{A_u\mu_0\omega_w d^2MN}{4\pi}i^{-\ell_u}\sigma_{q_{0}}\frac{e^{i2k_wr^{q_{0}}_t}}{{(r^{q_{0}}_t)}^2}e^{i\ell_u\varphi^{q_{0}}_t}k_wR\cos\theta^{q_{0}}_t\cdot\big(\frac{1}{2}{J_0}(k_wR\sin\theta^{q_{0}}_t){J_{\ell_u-1}}(k_wR\sin\theta^{q_{0}}_t)\nonumber\\
&-\frac{1}{2}{J_0}(k_wR\sin\theta^{q_{0}}_t){J_{\ell_u+1}}(k_wR\sin\theta^{q_{0}}_t)-{J_{\ell_u}}(k_wR\sin\theta^{q_{0}}_t){J_{1}}(k_wR\sin\theta^{q_{0}}_t)\big)s(\ell_u,k_w),
\end{align*}
\begin{align*}
\frac{\partial p(\ell_u,k_w,\boldsymbol{\vartheta})}{\partial \varphi^{q_{0}}_t}=&-j\frac{A_u\mu_0\omega_w d^2MN}{4\pi}i^{-\ell_u+1}\ell_u\sigma_{q_{0}}\frac{e^{i2k_wr^{q_{0}}_t}}{{(r^{q_{0}}_t)}^2}e^{i\ell_u\varphi^{q_{0}}_t}{J_{\ell_u}}(k_wR\sin\theta^{q_{0}}_t){J_0}(k_wR\sin\theta^{q_{0}}_t)s(\ell_u,k_w),
\end{align*}
\begin{align}
\frac{\partial p(\ell_u,k_w,\boldsymbol{\vartheta})}{\partial \Omega^{q_{_V}}_t}=&-j\frac{A_u\mu_0\omega_w d^2MN}{4\pi}i^{-\ell_u+1}\ell_ug_t^qt\sigma_{q_{_V}}\frac{e^{i2k_wr^{q_{_V}}_t}}{{(r^{q_{_V}}_t)}^2}e^{i\ell_u\varphi^{q_{_V}}_t}{J_{\ell_u}}(k_wR\sin\theta^{q_{_V}}_t){J_0}(k_wR\sin\theta^{q_{_V}}_t)\nonumber\\
&\ \cos(\Omega^{q_{_V}}_t t+\psi^{q_{_V}}_0+\delta^{q_{0}}_t), \qquad \qquad \qquad \qquad \qquad \qquad \qquad \qquad \qquad \qquad q=1,2,\cdots,Q,
\end{align}
\setcounter{equation}{30}%
\hrulefill
\end{figure*}

Thus, OAM-based rotating target imaging is completed. In the numerous estimated parameters, the 3-D estimated positions $\{(\hat{r}^{q_{0}}_t,\hat{\theta}^{q_{0}}_t,\hat{\varphi}^{q_{0}}_t)|q=1,2,\cdots,Q\}$ of the centroid of each target and the rotation velocities $\{\hat{\Omega}^{q_{V}}_t|q=1,2,\cdots,Q\}$ of the vertex of each target are regarded as the most important estimated parameters, which represent the key state characteristic of each target at time instant $t$.

\vspace{0.1cm}
\section{Beam Optimization of Joint OAM Radar-Communication System}
\vspace{0.1cm}

In the following part, we will first analyze the performances of the proposed radar-centric joint OAM RadCom system, and then optimize the transmitted integrated OAM beams to achieve the best tradeoff between the imaging and communication performances of the system.

\vspace{-0.0cm}
\subsection{PCRB of OAM-based Target Imaging}
\vspace{-0.0cm}
In this subsection, we discuss the PCRB of key state parameters $\{(r^{q_{0}}_t,\theta^{q_{0}}_t,\varphi^{q_{0}}_t,\Omega^{q_{_V}}_t)|q$$=$$1,2,\cdots,Q\}$ of $Q$ targets, which is regarded as the important metric for imaging performance \cite{Tichavsky1998Posterior}. First, all the key state parameters to be estimated are collected in $\boldsymbol{\vartheta}$ $=$ $[(r^{1_{0}}_t,$$\theta^{1_0}_t,$$\varphi^{1_{0}}_t,$$\Omega^{1_{V}}_t),$$\cdots,$$(r^{Q_{0}}_t,$$\theta^{Q_{0}}_t,$$\varphi^{Q_{0}}_t,$$\Omega^{Q_{_V}}_t)]^{\mathrm{T}}$. To derive the PCRB, the OAM echo signal received on the $u$ mode at the $w$ subcarrier in \eqref{Er} is rewritten as
\begin{align}\label{CRLB1}
\setcounter{equation}{32}
E_R(\ell_u,k_w)=p(\ell_u,k_w,\boldsymbol{\vartheta})+n(\ell_u,k_w),
\end{align}
where
\begin{align*}
&p(\ell_u,k_w,\boldsymbol{\vartheta})=-j\frac{A_u\mu_0\omega_w d^2MN}{4\pi}i^{-\ell_u}s(\ell_u,k_w)\times\nonumber\\
&\sum_{q=1}^{Q}\sum_{i=1}^{P}\sigma_{q_{i}} \frac{e^{i2k_wr^{q_{i}}_t}}{{(r^{q_{i}}_t)}^2}e^{i\ell_u\varphi^{q_{i}}_t}{J_{\ell_u}}(k_wR\sin\theta^{q_{i}}_t){J_0}(k_wR\sin\theta^{q_{i}}_t).
\end{align*}
%is a random signal with respect to $\boldsymbol{\vartheta}$.
Then, all the OAM echo signals received on the $U$ modes at the $W$ subcarriers can be expressed in compact form as
\begin{align}\label{CRLB2}
\mathbf{E}_R=\mathbf{P}(\boldsymbol{\vartheta})+\mathbf{N},
\end{align}
where $\mathbf{E}_R=[E_R(\ell_u,k_w)]_{U\times W}$, $\mathbf{P}(\boldsymbol{\vartheta})=[p(\ell_u,k_w,\boldsymbol{\vartheta})]_{U\times W}$, and $\mathbf{N}=[n(\ell_u,k_w)]_{U\times W}$.

After that, the Fisher information matrix $\mathbf{J}$ \cite{Tichavsky1998Posterior} with respect to $\boldsymbol{\vartheta}$ can be expressed as \eqref{Fisher}, whose detailed derivation is given in Appendix A. Based on \eqref{Fisher}, the PCRB of the key state parameter to be estimated in $\boldsymbol{\vartheta}$ can be expressed as
\begin{align}\label{PCRB}
\setcounter{equation}{35}
\mathbb{E}\left\{(\vartheta_i-\hat{\vartheta}_i)^2\right\}\geq[\mathbf{J}^{-1}]_{ii},
\end{align}
where $\vartheta_i$ is the $i$-th key state parameter in $\boldsymbol{\vartheta}$, $\hat{\vartheta}_i$ is the estimate of $\vartheta_i$, and $[\mathbf{J}^{-1}]_{ii}$ is the $i$-th diagonal element of the inverse matrix of $\mathbf{J}$.
%From \eqref{Fisher} and \eqref{PCRB}, we can observe that the PCRB of all the state parameters to be estimated can be optimized by designing the weights $\{A_u|u=1,2,\cdots,U\}$ of integrated OAM beams, which will be discussed in detail in Section V.

\vspace{-0.2cm}
\subsection{Data Rate of Joint OAM RadCom System}

In this subsection, we discuss the transmission rate of the radar-centric joint OAM RadCom system based on \eqref{y}. During a period, the equivalent baseband signal vector $\mathbf{y}_{\bar{q}}(k_w)$ received by the specific target at the $w$-th subcarrier can be expressed as
\begin{align} \label{vectory}
\mathbf{y}_{\bar{q}}(k_w)&=\mathbf{H}_{\bar{q}}(k_w)\mathbf{X}(k_w)+\mathbf{z}_{\bar{q}}(k_w)\nonumber\\
&=\mathbf{H}_{\bar{q}}(k_w)\mathbf{F}\mathbf{A}\mathbf{S}(k_w)+\mathbf{z}_{\bar{q}}(k_w),
\end{align}
where $\mathbf{H}_{\bar{q}}(k_w)$ $=$ $[h^{\bar{q}}_m(k_w)]_{1\times M}$ is the channel matrix from the dual-function transmitter to the specific target, $\mathbf{X}(k_w)$ $=$ $[A_u e^{i\ell_u\varphi_m}s(\ell_u,k_w)]_{M\times U}$ is the transmit signal matrix generated by the dual-function transmitter, $\mathbf{F}$ $=$ $[e^{i\ell_u\varphi_m}]_{M\times U}$ is the right circularly shifted (partial) inverse fast Fourier transform (IFFT) matrix used to generate integrated OAM beams, $\mathbf{A}$ $=$ $\textrm{diag}\{A_1,\cdots,A_u\}$ is the $U\!$-dimensional weight matrix of the integrated OAM beams, $\mathbf{S}(k_w)=\textrm{diag}\{s(\ell_1,k_w),\cdots,s(\ell_U,k_w)\}$ is the $U\!$-dimensional integrated OAM symbol matrix at the $w$-th subcarrier, and $\mathbf{z}_{\bar{q}}(k_w)=[z(\ell_1,k_w),\cdots,z(\ell_U,k_w)]$ is the noise vector.

%radar-centric
In the proposed radar-centric joint OAM RadCom system, we assume to detect the received OAM signals by employing the simplest zero-forcing detection method. Then, after obtaining the channel state information (CSI) by the parametric channel estimation method \cite{Chen2020Multi,Long2021Joint,Long2021AoA}, the detected data symbol matrix at the specific target can be expressed as
\begin{align} \label{detectedy}
\mathbf{\hat{S}}_{\bar{q}}(k_w)=&(\mathbf{\hat{H}}_{\bar{q}}(k_w)\mathbf{F})^\dagger\big(\mathbf{H}_{\bar{q}}(k_w)\mathbf{F}\mathbf{A}\mathbf{S}(k_w)+\mathbf{z}_{\bar{q}}(k_w)\big)\nonumber\\
=&\mathbf{A}\mathbf{S}(k_w)+\bm{\mathcal{I}}_{\bar{q}}(k_w)+\mathbf{Z}'_{\bar{q}}(k_w),
\end{align}
where $\mathbf{\hat{H}}_{\bar{q}}(k_w)$ is the estimated channel matrix, $\bm{\mathcal{I}}_{\bar{q}}(k_w)=\big((\mathbf{\hat{H}}_{\bar{q}}(k_w)\mathbf{F})^\dagger\mathbf{H}_{\bar{q}}(k_w)\mathbf{F}-\mathbf{I}_U\big)\mathbf{A}\mathbf{S}(k_w)$ is the remaining interferences induced by the error in $\mathbf{\hat{H}}_{\bar{q}}(k_w)$, $\mathbf{I}_U$ is the $U\!$-dimensional unit matrix, $\mathbf{Z}'_{\bar{q}}(k_w)=(\mathbf{\hat{H}}_{\bar{q}}(k_w)\mathbf{F})^\dagger\mathbf{z}_{\bar{q}}(k_w)$.

\begin{figure}[t] %figure1
\setlength{\abovecaptionskip}{0.0cm}   %è°æŽåŸçæ é¢äžåŸè·çŠ»
\setlength{\belowcaptionskip}{-0.2cm}   %è°æŽåŸçæ é¢äžäžæè·çŠ»
\footnotesize
\begin{center}
\includegraphics[width=7.0cm,height=10.2cm]{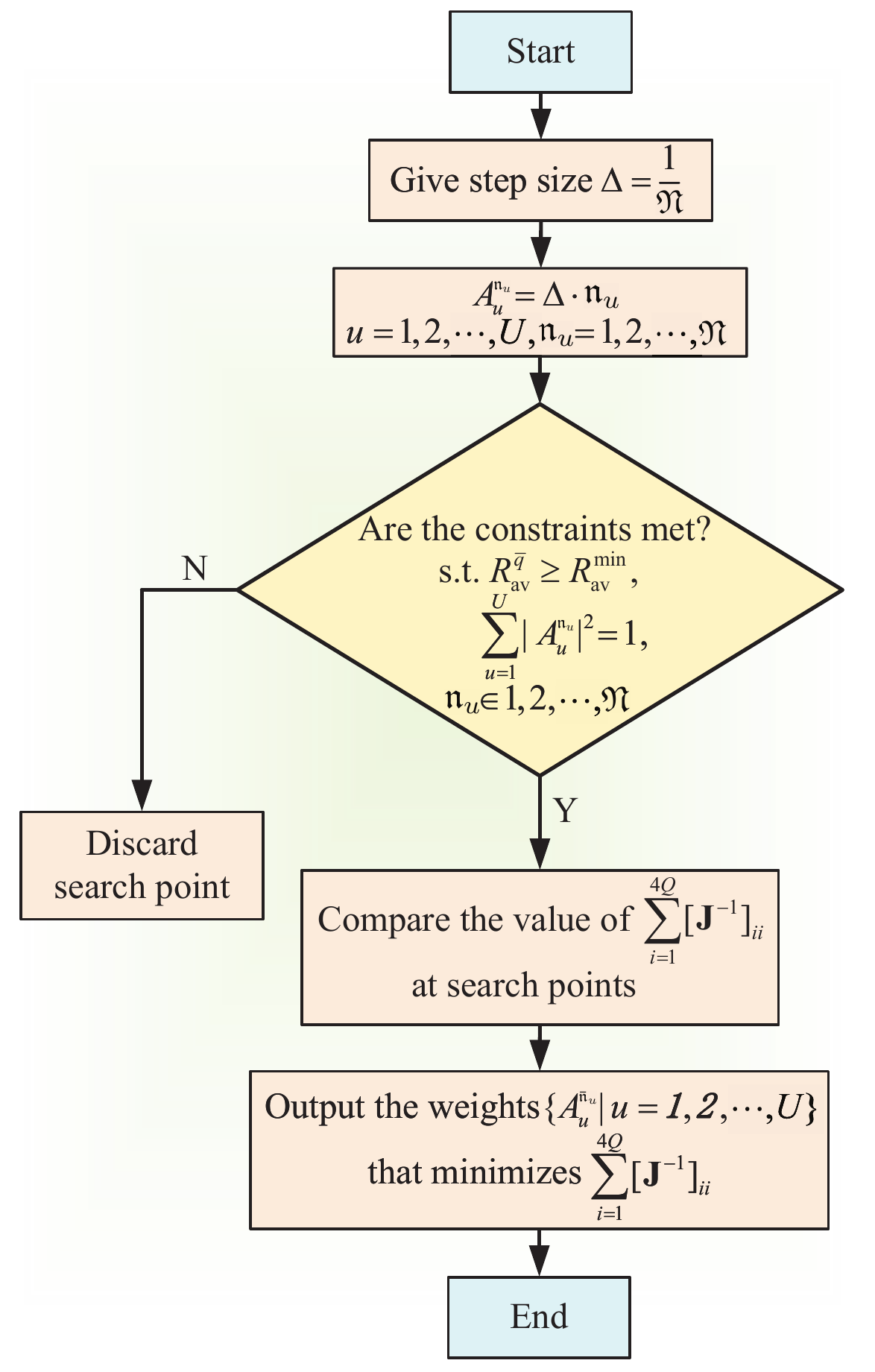}
\end{center}
\caption{The flowchart of the weight optimization of integrated OAM beams with the exhaustive search method.}
\label{Fig5}
\end{figure}

Define $\mathbf{R}^{\bar{q}}_{\mathbf{A}\mathbf{S}}(k_w)$, $\mathbf{R}^{\bar{q}}_{\bm{\mathcal{I}}}(k_w)$ and $\mathbf{R}^{\bar{q}}_{\mathbf{Z}'}(k_w)$ as the $U\times U$ covariance matrices of $\mathbf{A}\mathbf{S}(k_w)$, $\bm{\mathcal{I}}_{\bar{q}}(k_w)$ and $\mathbf{Z}'_{\bar{q}}(k_w)$, i.e.,
\begin{align} \label{CovarianceMatrix}
&\quad\quad\quad\quad\quad \mathbf{R}^{\bar{q}}_{\mathbf{A}\mathbf{S}}(k_w)=\mathbf{A} \mathbb{E}\left\{\mathbf{S}(k_w)\mathbf{S}^\mathrm{H}(k_w)\right\}\mathbf{A}^\mathrm{H},\nonumber\\
&\mathbf{R}^{\bar{q}}_{\bm{\mathcal{I}}}(k_w)=\big((\mathbf{\hat{H}}_{\bar{q}}(k_w)\mathbf{F})^\dagger\mathbf{H}_{\bar{q}}(k_w)\mathbf{F}\!-\!\mathbf{I}_U\big)\mathbf{A}\mathbb{E}\left\{\mathbf{S}(k_w)\mathbf{S}^\mathrm{H}(k_w)\!\right\}\nonumber\\
&\qquad\qquad\quad \mathbf{A}^\mathrm{H}\big((\mathbf{\hat{H}}_{\bar{q}}(k_w)\mathbf{F})^\dagger\mathbf{H}_{\bar{q}}(k_w)\mathbf{F}-\mathbf{I}_U\big)^\mathrm{H},\nonumber\\
&\mathbf{R}^{\bar{q}}_{\mathbf{Z}'}(k_w)=(\mathbf{\hat{H}}_{\bar{q}}(k_w)\mathbf{F})^\dagger\mathbb{E}\left\{\mathbf{z}_{\bar{q}}(k_w)\mathbf{z}_{\bar{q}}^\mathrm{H}(k_w)\right\}\big((\mathbf{\hat{H}}_{\bar{q}}(k_w)\mathbf{F})^\dagger\big)^\mathrm{H},
\end{align}
then, the signal-to-interference-plus-noise ratio (SINR) on the $u$-th OAM mode at the $w$-th subcarrier can be formulated as
\begin{equation} \label{SINR}
\textrm{SINR}_{\bar{q}}(\ell_u,k_w)=\frac{\left[\mathbf{R}^{\bar{q}}_{\mathbf{A}\mathbf{S}}(k_w)\right] _{uu}}{\left[\mathbf{R}^{\bar{q}}_{\bm{\mathcal{I}}}(k_w)\right] _{uu}+ \left[\mathbf{R}^{\bar{q}}_{\mathbf{Z}'}(k_w)\right]_{uu}},
\end{equation}
where $[$ $\cdot$ $]_{uu}$ is the $u$-th diagonal element of the matrix. Therefore, the average data rate at the specific communication target during each period can be written as
%
%\begin{align} \label{AR}
%R^{\bar{q}}_{\textrm{av}} =&\frac{1}{W}\frac{1}{U}\sum_{w=1}^W\sum_{u=1}^U
%\log_2\left(1 + \textrm{SINR}_{\bar{q}}(\ell_u,k_w)\right).
%\end{align}
%

\begin{align} \label{AR}
R^{\bar{q}}_{\textrm{av}} =\frac{1}{U}\sum_{w=1}^W\sum_{u=1}^U
\log_2\left(1 + \textrm{SINR}_{\bar{q}}(\ell_u,k_w)\right).
\end{align}
%From \eqref{AR}, we can observe that the average data rate of the integrated OAM RadCom system is closely associated with the weights $\{A_u|u=1,2,\cdots,U\}$ of integrated OAM beams, whose optimization will be discussed in detail in the following part.

\subsection{Optimization of Integrated OAM Beams}

From \eqref{PCRB} and \eqref{AR}, we can see that both the imaging and communication performances of the joint OAM RadCom system are closely associated with the weights $\{A_u|u=1,2,\cdots,U\}$ of integrated OAM beams. Different power distribution schemes will directly affect the performances of the system. To achieve the best tradeoff between the imaging and communication performances, we formulate the optimization problem to minimize the PCRB, subject to the data rate constraint for the specific target as well as a transmit power budget, i.e.,
\begin{align} \label{optimization}
&\min_{_{A_1,\cdots,A_U}}\ \sum_{i=1}^{4Q}\ [\mathbf{J}^{-1}]_{ii}\nonumber\\
&\textrm{s.t.} \ R^{\bar{q}}_{\textrm{av}}\geq R^{\textrm{min}}_{\textrm{av}},\nonumber\\
&\quad \ \sum_{u=1}^{U}|A_u|^2=1,
\end{align}
where $R^{\textrm{min}}_{\textrm{av}}$ is the minimum average data rate required by the specific target.
%Due to the complexity of $[\mathbf{J}^{-1}]_{ii}$ and $R^{\bar{q}}_{\textrm{av}}$, the optimal value of the problem \eqref{optimization} is not easily obtained by the common optimization methods, such as the Lagrange multiplier method.
%Therefore,
%Considering the complexity of $[\mathbf{J}^{-1}]_{ii}$ and $R^{\bar{q}}_{\textrm{av}}$,

%
\begin{figure*}[t]
\setlength{\abovecaptionskip}{-0.2cm}   %è°æŽåŸçæ é¢äžåŸè·çŠ»
\setlength{\belowcaptionskip}{-0.2cm}   %è°æŽåŸçæ é¢äžäžæè·çŠ»
\begin{center}
\includegraphics[scale=0.70]{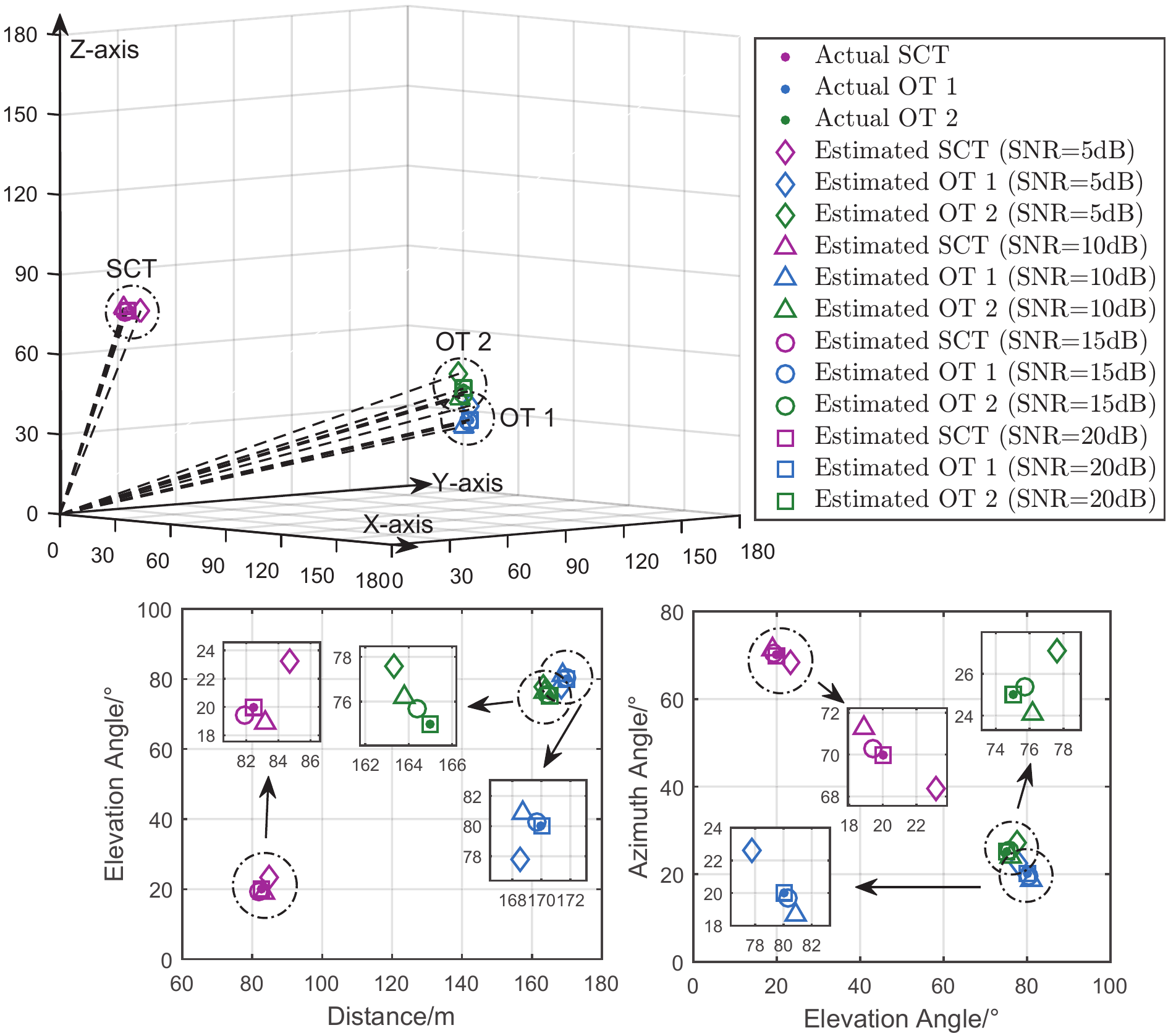}
\end{center}
\caption{The 3-D position estimation results of the proposed method. SCT: the specific communication target, OT: the other target.}
\label{Fig6}
\end{figure*}
\begin{figure*}[t]
\setlength{\abovecaptionskip}{-0.1cm}   %è°æŽåŸçæ é¢äžåŸè·çŠ»
\setlength{\belowcaptionskip}{-0.2cm}   %è°æŽåŸçæ é¢äžäžæè·çŠ»
\centering
\subfigure[]{
\includegraphics[scale=0.345]{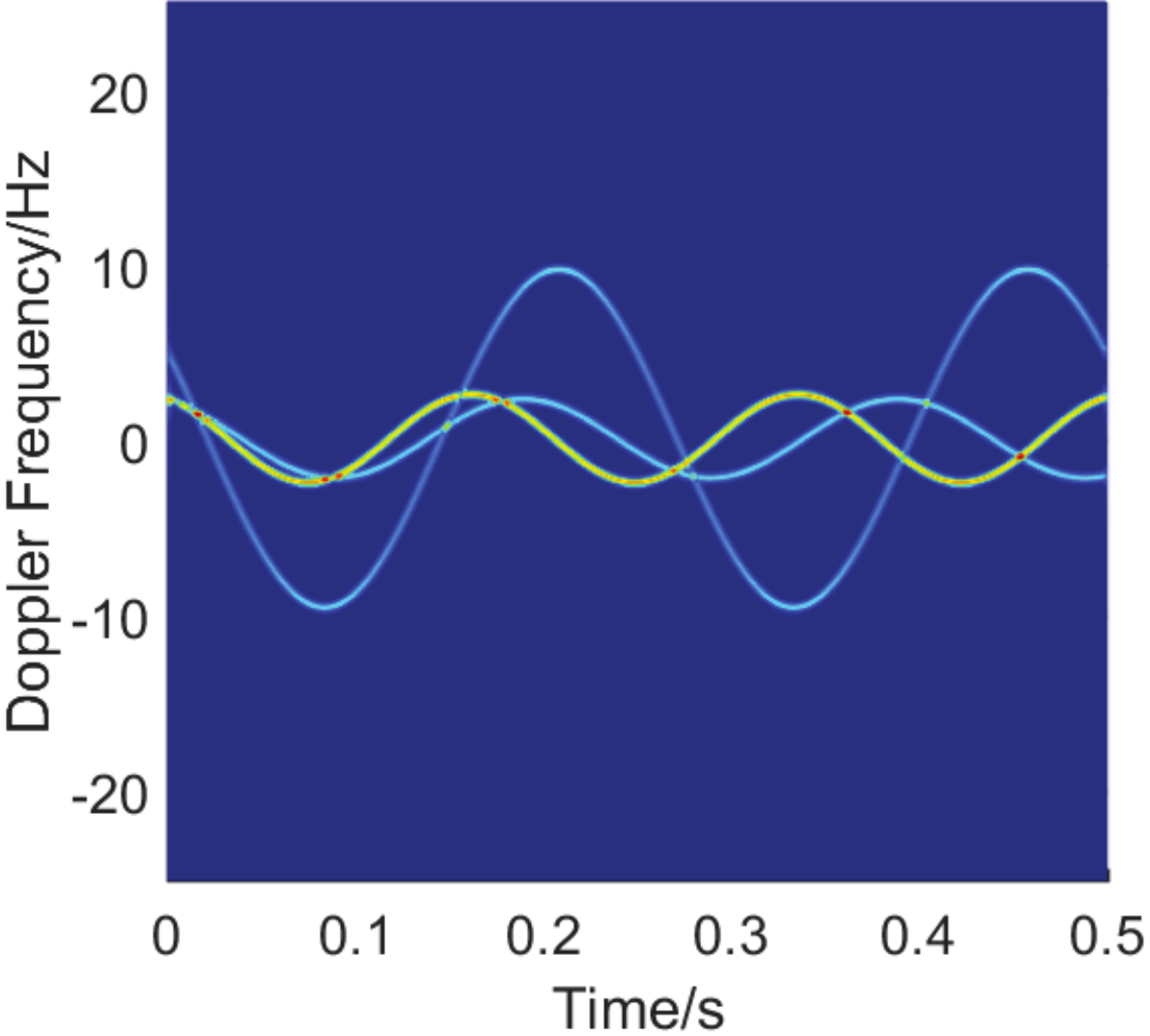}
}
\subfigure[]{
\includegraphics[scale=0.347]{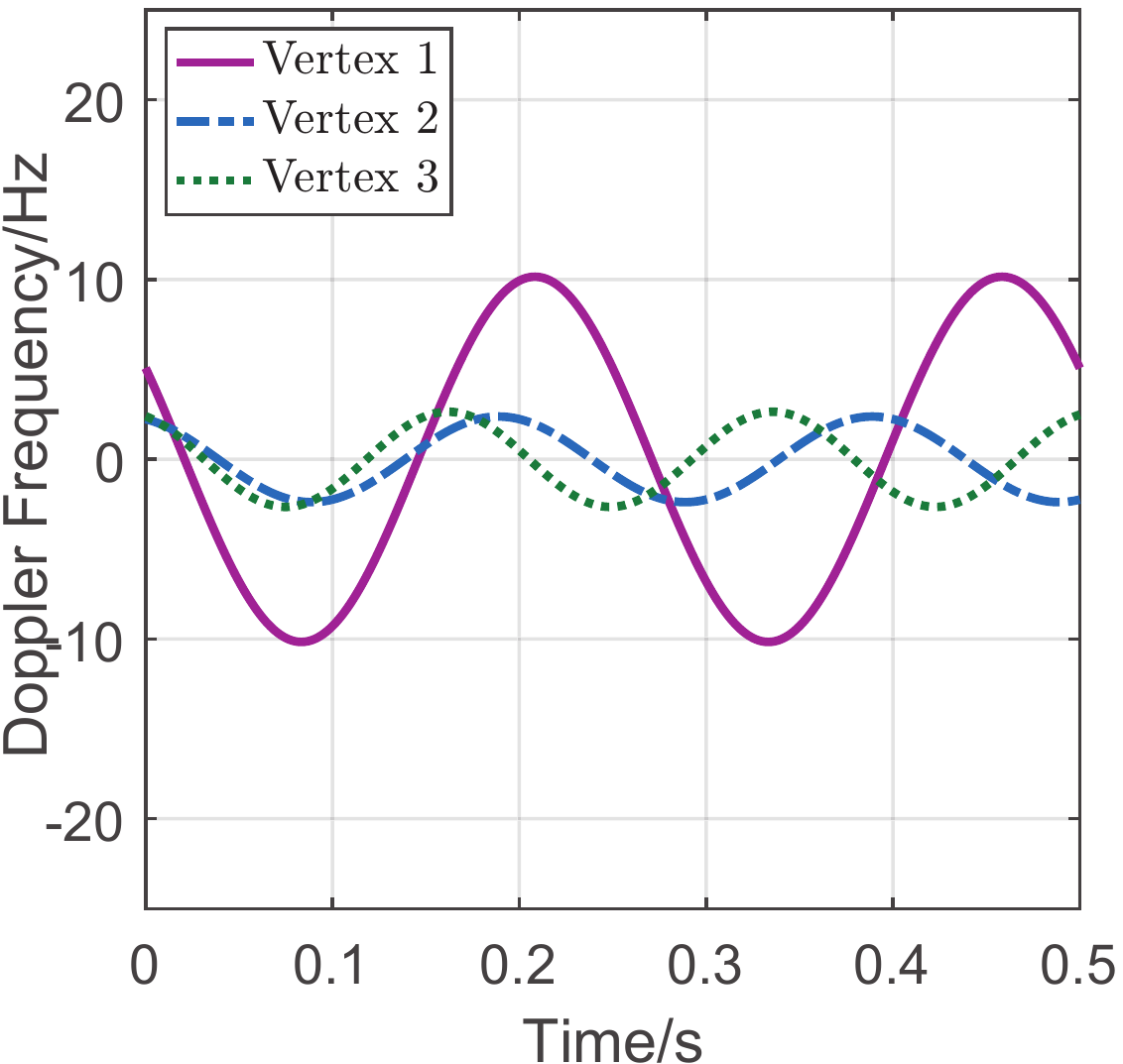}
}
\subfigure[]{
\includegraphics[scale=0.3465]{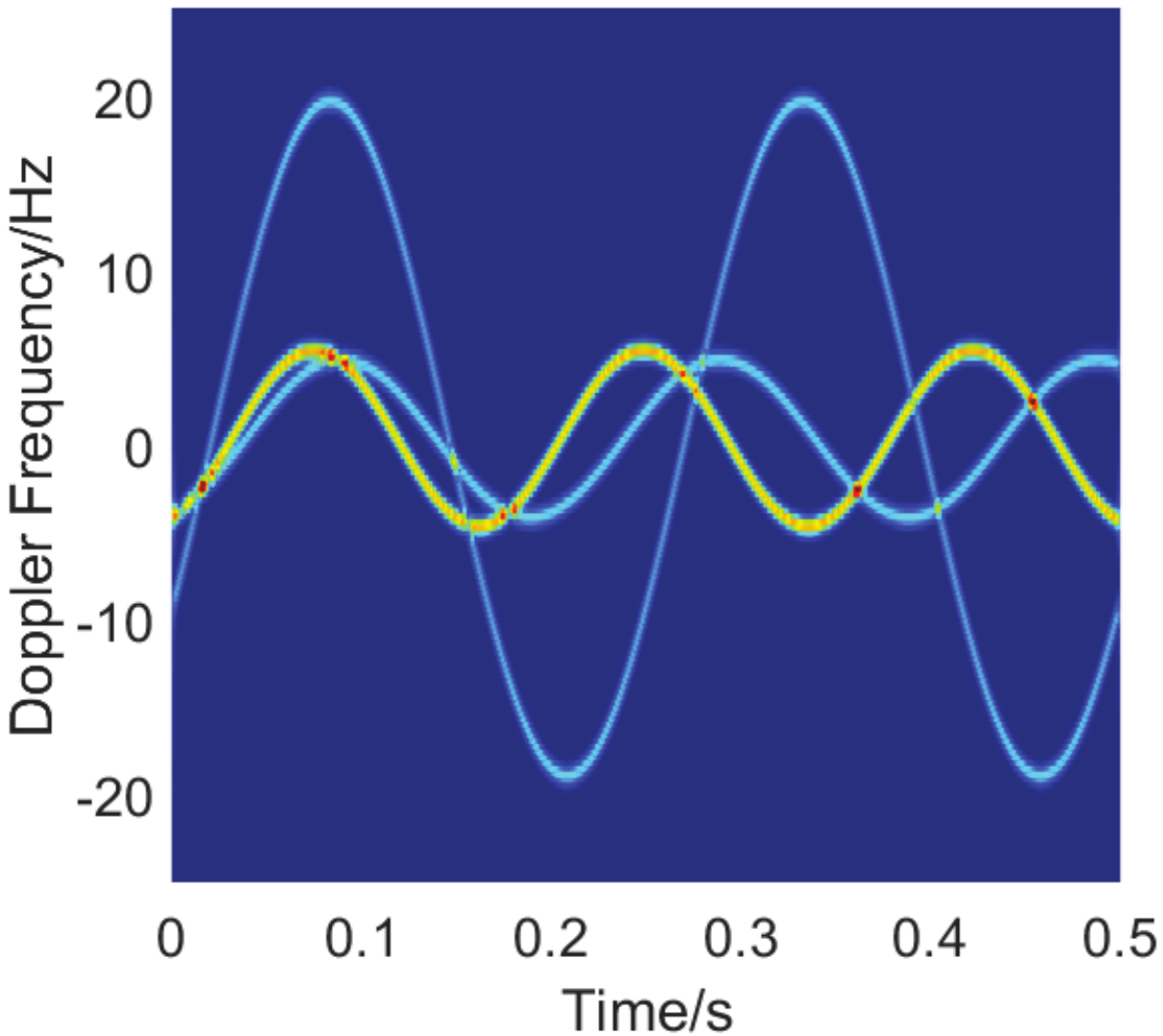}
}
\subfigure[]{
\includegraphics[scale=0.3460]{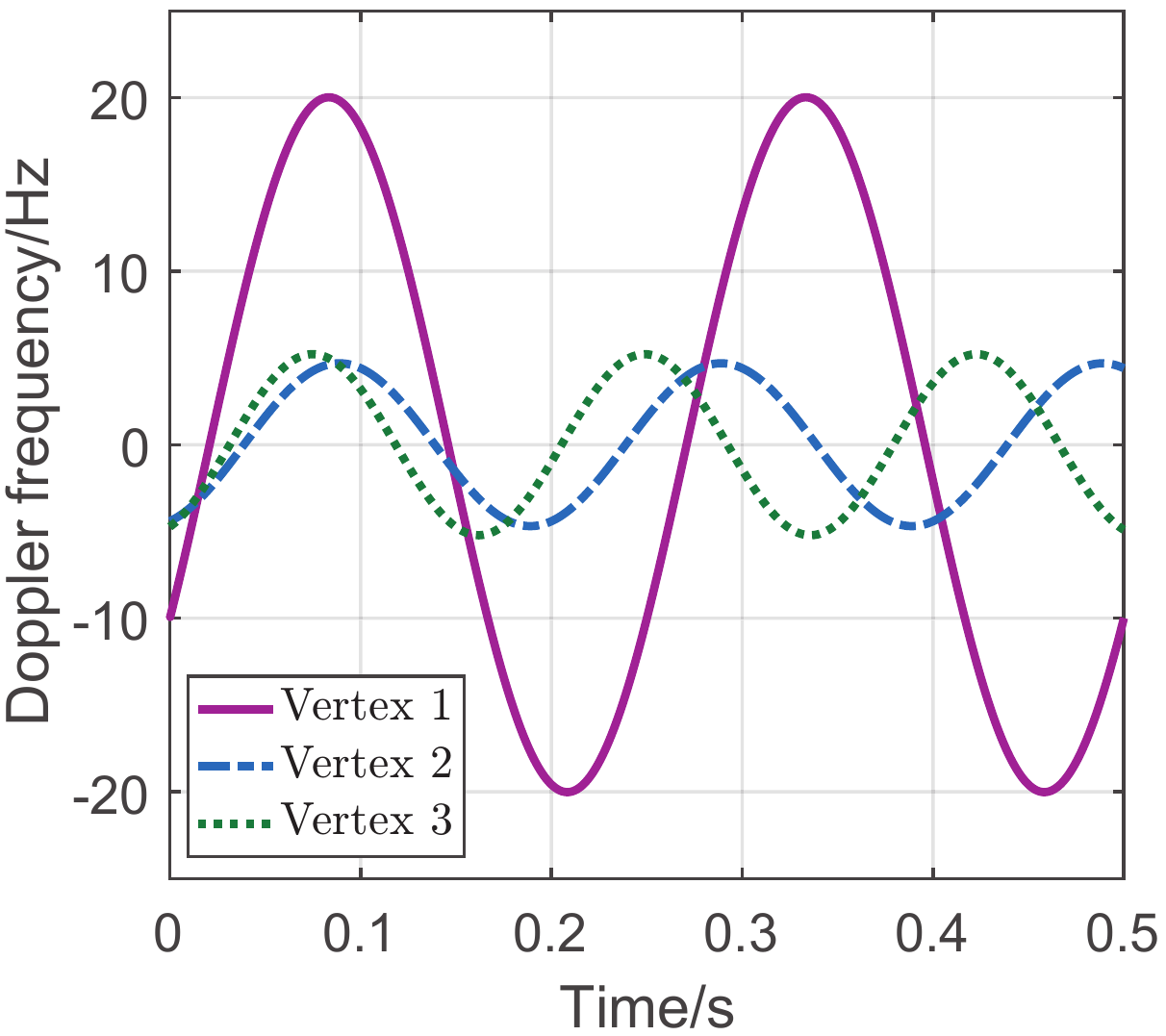}
}
\caption{The time-frequency simulation results at $15$dB. (a) STFT image with $\ell_u=-1$ . (b) Theoretical Doppler curve with $\ell_u=-1$. (c) STFT image with $\ell_u=+2$ . (d) Theoretical Doppler curve with $\ell_u=+2$.}
\label{Fig7}
\end{figure*}
\begin{figure*}[t]
\setlength{\abovecaptionskip}{-0.1cm}   %è°æŽåŸçæ é¢äžåŸè·çŠ»
\setlength{\belowcaptionskip}{-0.2cm}   %è°æŽåŸçæ é¢äžäžæè·çŠ»
\centering
\subfigure[]{
\includegraphics[scale=0.35]{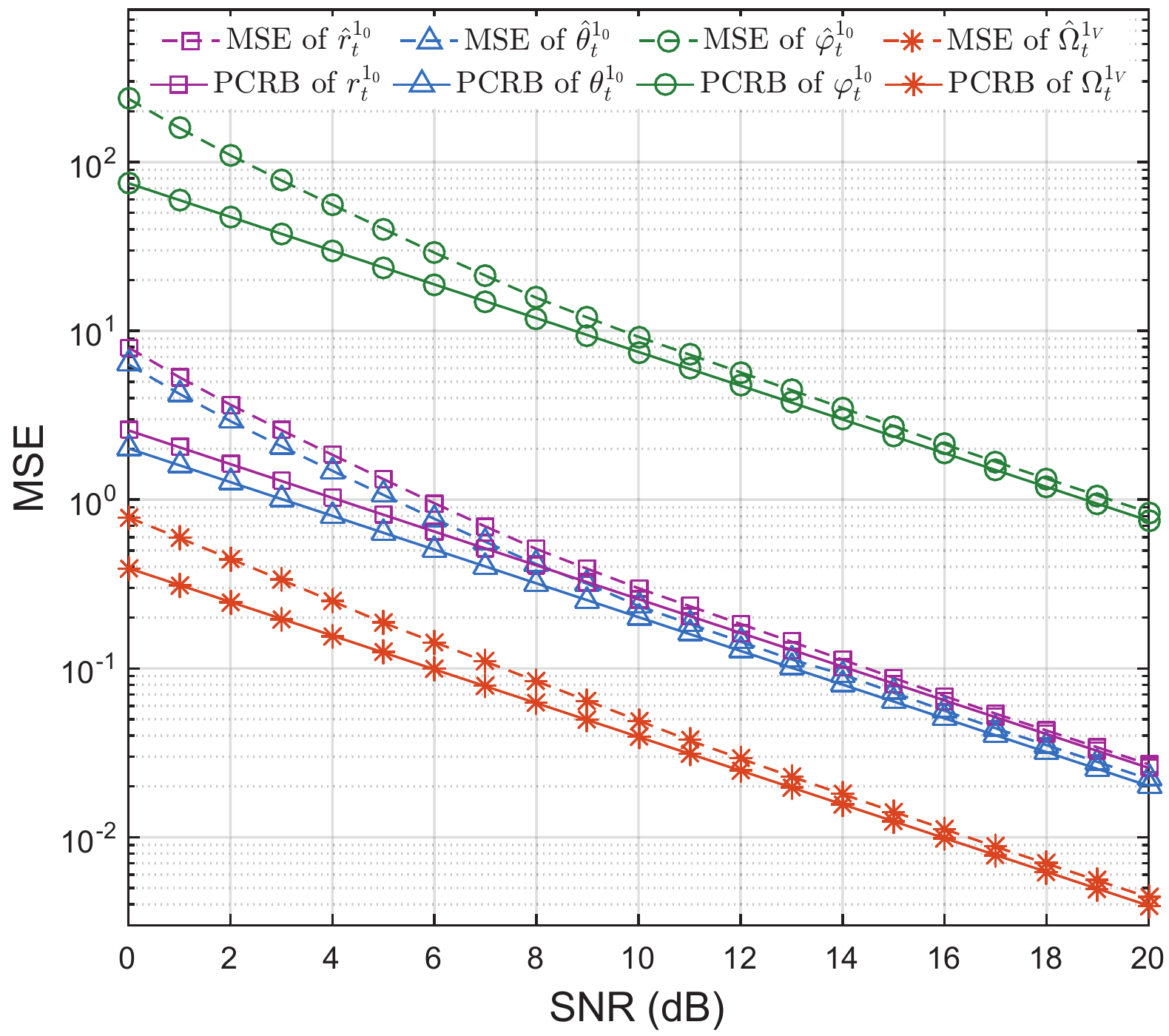}
}
\subfigure[]{
\includegraphics[scale=0.348]{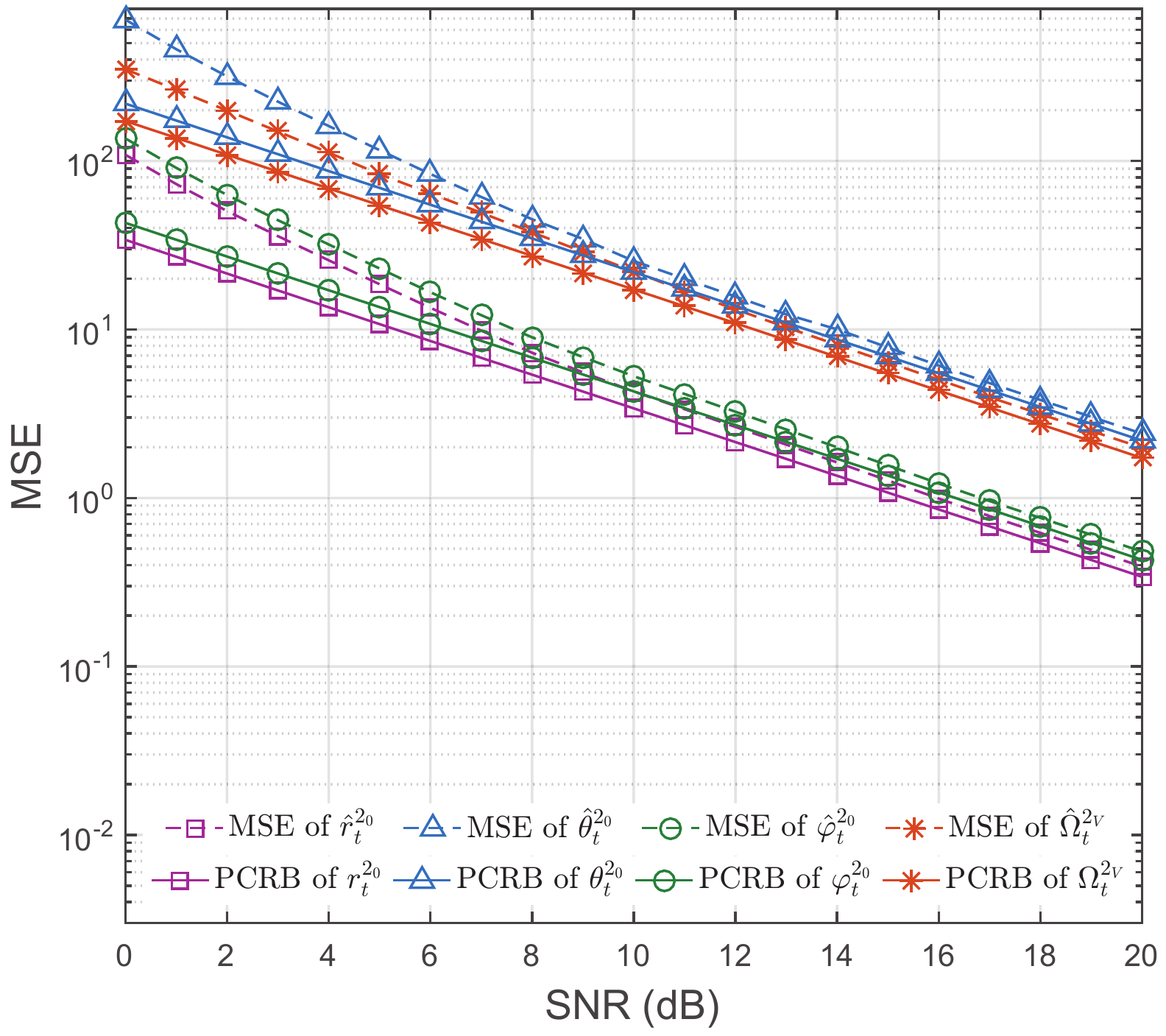}
}
\subfigure[]{
\includegraphics[scale=0.347]{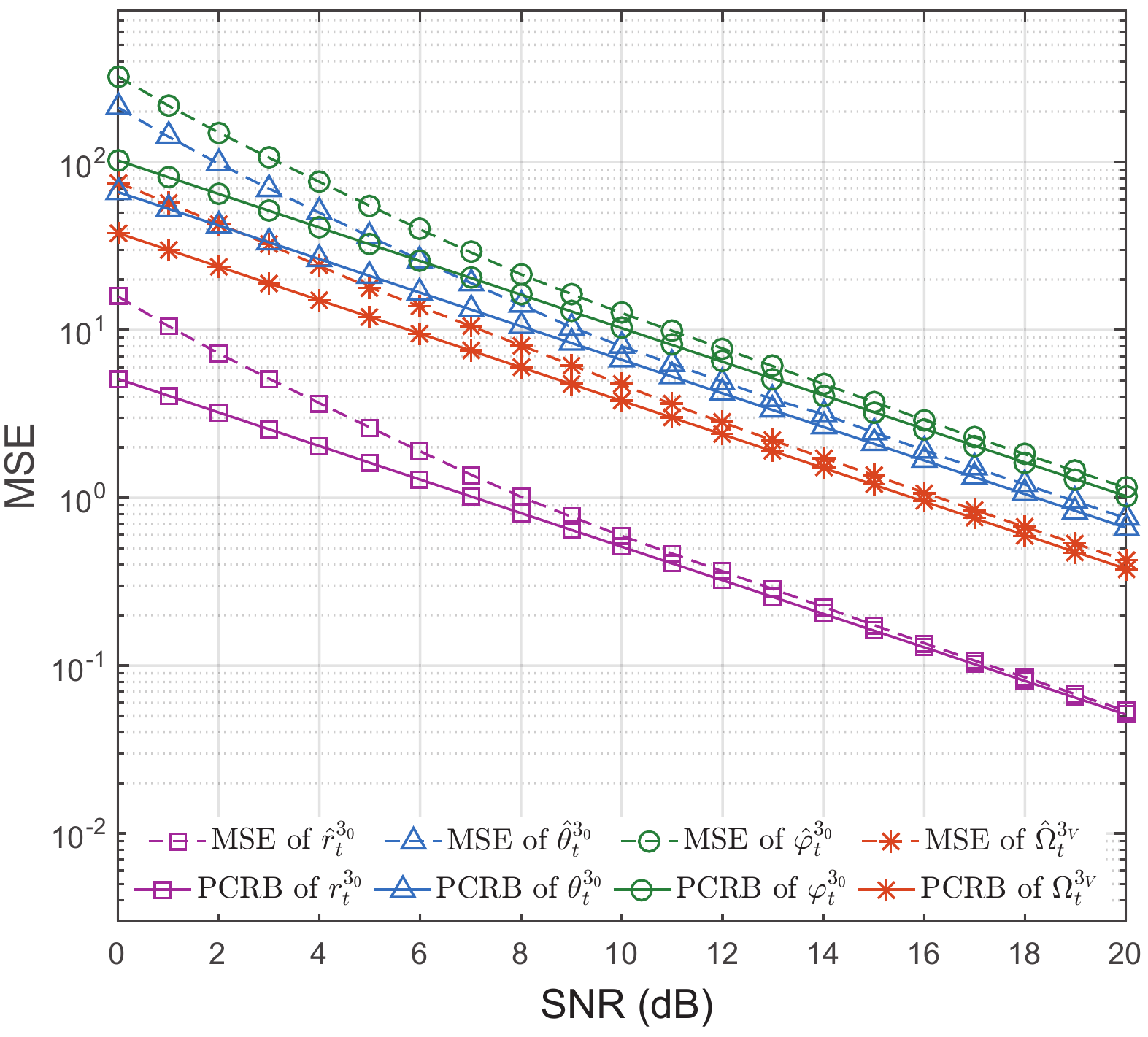}
}
\caption{The comparison of the MSEs and PCRBs. (a) Specific communication target $(\hat{r}^{1_0}_t,\hat{\theta}^{1_0}_t,\hat{\varphi}^{1_0}_t,\hat{\Omega}^{1_V}_t)$. (b) Other target 1 $(\hat{r}^{2_0}_t,\hat{\theta}^{2_0}_t,\hat{\varphi}^{2_0}_t,\hat{\Omega}^{2_V}_t)$. (c) Other target 2 $(\hat{r}^{3_0}_t,\hat{\theta}^{3_0}_t,\hat{\varphi}^{3_0}_t,\hat{\Omega}^{3_V}_t)$.}
\label{addfig}
\end{figure*}
\begin{figure*}[t]
\setlength{\abovecaptionskip}{-0.1cm}   %è°æŽåŸçæ é¢äžåŸè·çŠ»
\setlength{\belowcaptionskip}{-0.4cm}   %è°æŽåŸçæ é¢äžäžæè·çŠ»
\centering
\subfigure[]{
\includegraphics[scale=0.435]{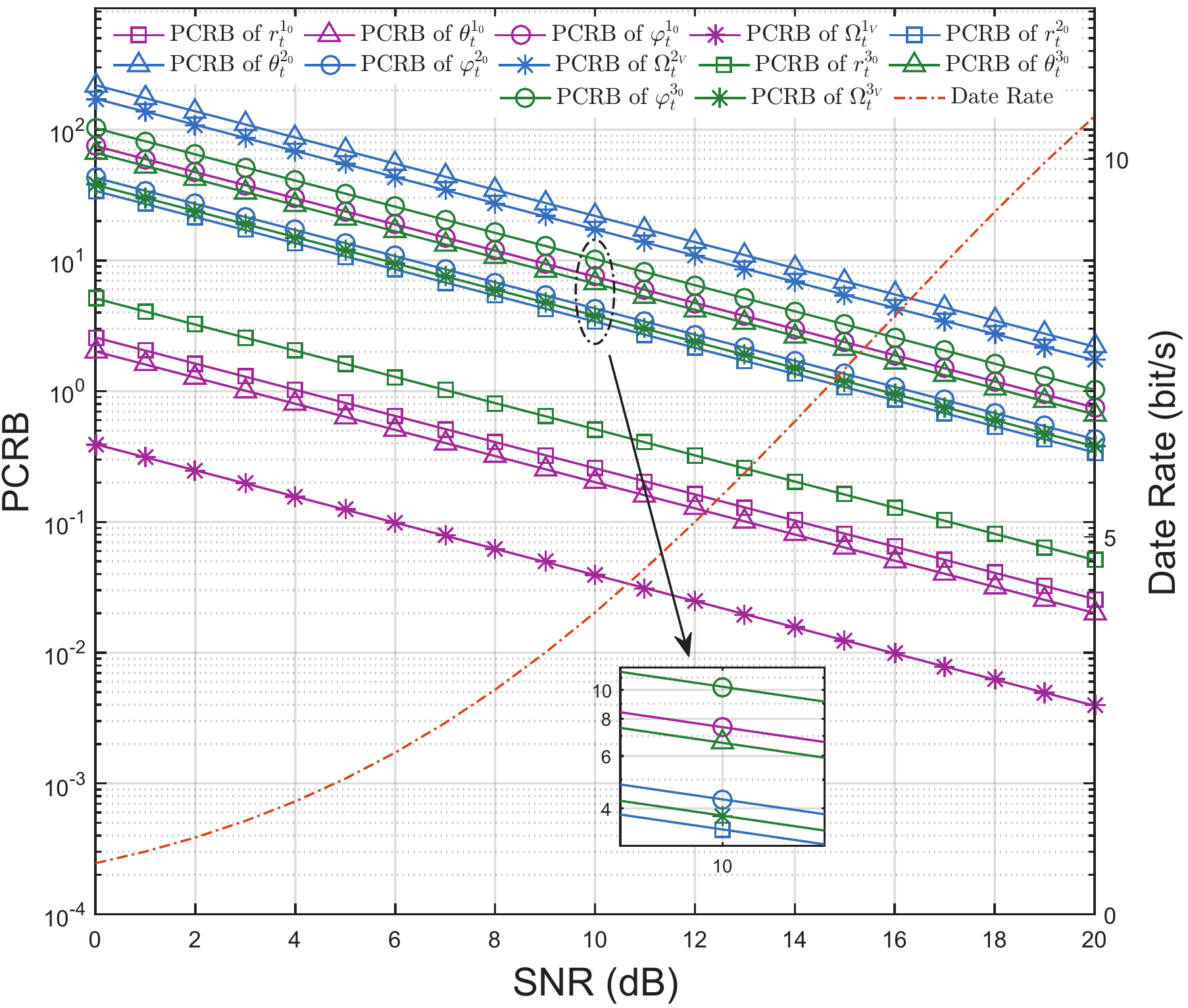}
}
\subfigure[]{
\includegraphics[scale=0.440]{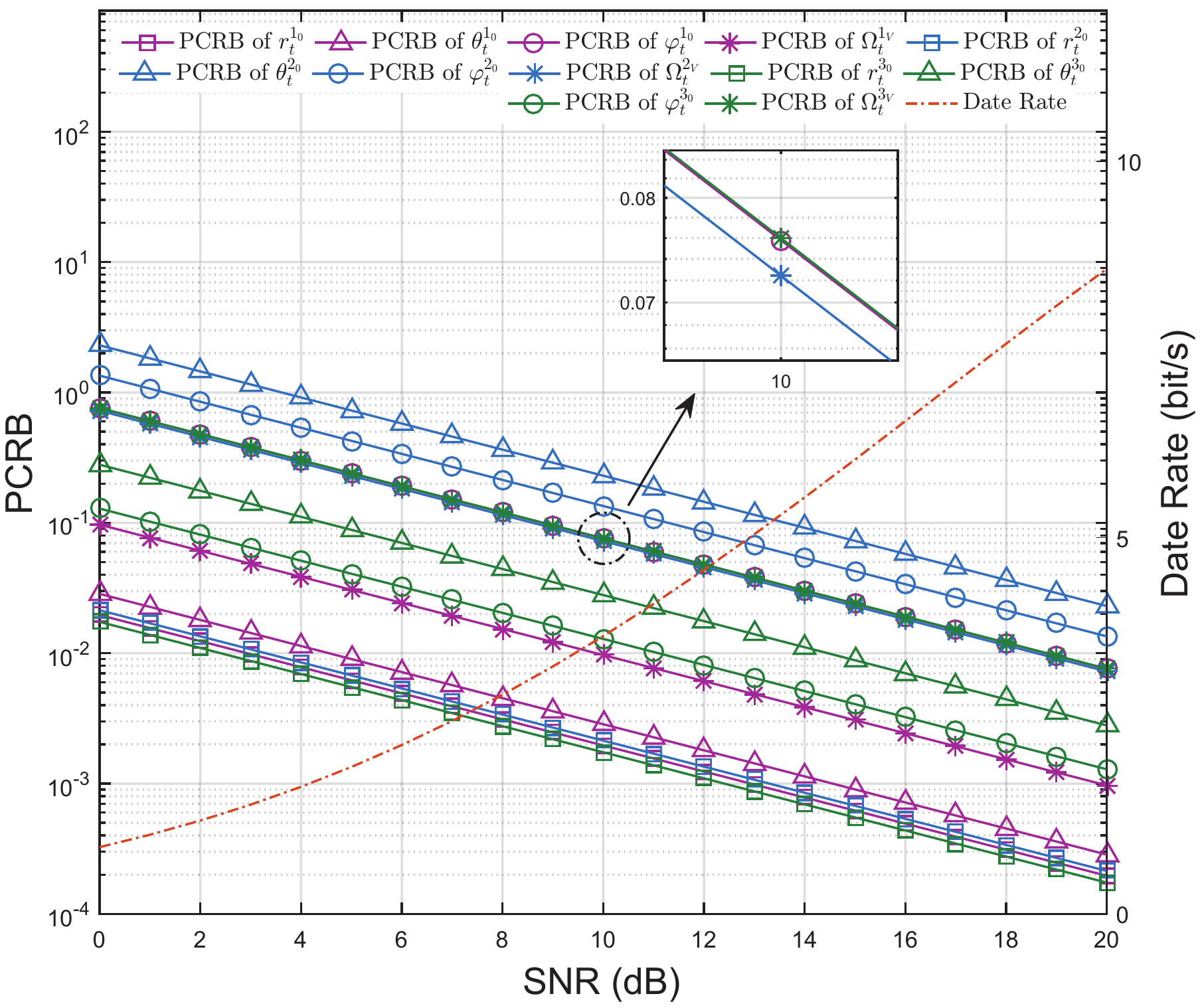}
}
\caption{The PCRB and data rate of the radar-centric joint OAM RadCom system. (a) Average power distribution. (b) Random power distribution.}
\label{Fig8}
\end{figure*}

Since $\sum_{i=1}^{4Q}\ [\mathbf{J}^{-1}]_{ii}$ can not be expressed as a function of $\boldsymbol{\vartheta}$ in the closed form, we propose to solve the optimization problem \eqref{optimization} by the exhaustive search method, which is the simplest optimization method \cite{NAYAK202071Single}. In the exhaustive search method, the optimal value of the problem \eqref{optimization} can be obtained by calculating the function values at several equally spaced points. We first select a search step size $\Delta=1/\mathfrak{N}$ for the weight $A_u\in (0,1)$, and calculate the function values of all search points $\{A_u^{\mathfrak{n}_u}|{\mathfrak{n}_u}=1,2,\cdots,\mathfrak{N}, u=1,2,\cdots,U\}$ based on the problem \eqref{optimization}, where $\mathfrak{N}$ is the number of intermediate points, $A_u^{\mathfrak{n}_u}$ is the $\mathfrak{n}_u$-th intermediate point of the weight $A_u$. According to the calculation result, the search point $\{A_u^{\bar{\mathfrak{n}}_u}|u=1,2,\cdots,U\}$ that minimizes the PCRB and meets the constraints is adopted as the optimal value of the problem \eqref{optimization}, and the detailed procedure is summarized in Fig.\ref{Fig5}. Based on the optimal weights, the dual-function transmitter distributes power for $U$ integrated OAM beams in a period to achieve the best performance tradeoff of the radar-centric joint OAM RadCom system.

\section{Numerical Simulations and Results}

In this section, we show the performances of the proposed scheme by numerical simulations. We first verify the proposed OAM-based 3-D position estimation and rotation velocity detection methods at different signal-to-noise ratios (SNRs), and compare the mean square error (MSE) of the proposed methods with the PCRB of the OAM-based imaging. Then,
%the effect of the transmit power on
we plot
the PCRB and data rate of the radar-centric joint OAM RadCom system
% is  exhibited.
 vs. the SNR.
Finally, we show the  performance tradeoff under the different data rate constraints. Unless otherwise stated, the SNRs in all the figures are defined as the ratio of the transmitted signal power versus the noise power.

We choose $W$ $=$ $16$ subcarriers from $9.979$GHz to $10.695$GHz corresponding to the wave numbers $k_1,$ $k_2,$ $\cdots,$ $k_{16}$ $=$ $209,$ $210,$ $\dots,$ $224$, $M$ $=$ $N$ $=$ $17$, $U$ $=$ $16$ OAM modes with $\ell_1,$ $\ell_2,$ $\cdots,$ $\ell_{16}$ $=$ $-8,$ $-7,$ $\cdots,$ $+7$, $R$ $=$ $30\lambda_1$, $\lambda_1$ $=$ $2\pi/k_1$, $Q$ $=$ $3$ with the key state parameters of the specific communication target $(r^{1_0}_t,$ $\theta^{1_0}_t,$ $\varphi^{1_0}_t, \Omega^{1_V}_{t})$ $=$ $(82.5\textrm{m},$ $20^{\circ},$ $70^{\circ},8\pi)$ and other targets $(r^{2_0}_t,$ $\theta^{2_0}_t,$ $\varphi^{2_0}_t,\Omega^{2_V}_{t})$ $=$ $(170\textrm{m},$ $80^{\circ},$ $20^{\circ},10\pi)$, $(r^{3_0}_t,$ $\theta^{3_0}_t,$ $\varphi^{3_0}_t,,\Omega^{3_V}_{t})$ $=$ $(165\textrm{m},$ $75^{\circ},$ $25^{\circ},11.5\pi)$ and $P = 3$ scattering points per target.

Then, by using the proposed method, the estimated positions of $Q$ targets are shown in Fig.\ref{Fig6}. As we can see from the figure,  the estimated positions of $Q$ targets approach the actual positions as the  SNR increases, e.g., when SNR reaches $20$dB, it is $(\hat{r}^{1_0}_t,$ $\hat{\theta}^{1_0}_t,$ $\hat{\varphi}^{1_0}_t)$ $=$ $(82.502\textrm{m},$ $19.997^{\circ},$ $69.995^{\circ})$, $(\hat{r}^{2_0}_t,$ $\hat{\theta}^{2_0}_t,$ $\hat{\varphi}^{2_0}_t)$ $=$ $(169.996\textrm{m},$ $80.002^{\circ},$ $19.998^{\circ})$, $(\hat{r}^{3_0}_t,$ $\hat{\theta}^{3_0}_t,$ $\hat{\varphi}^{3_0}_t)$ $=$ $(165.000\textrm{m},$ $74.996^{\circ},$ $24.994^{\circ})$, which are very close to the actual positions.

\begin{table}[t]
\vspace{0.1cm}
\caption{The estimation results of $\Omega^{q}_{t}$.}
\vspace{-0.4cm}
\begin{center}
\setlength{\tabcolsep}{5mm}{
\begin{tabular}{cccc}
  \toprule
                     &\textbf{$\hat{\Omega}^{1_V}_{t}$}       &\textbf{$\hat{\Omega}^{2_V}_{t}$}        &\textbf{$\hat{\Omega}^{3_V}_{t}$}\\
  \midrule
  {$5$dB}            &$8.281\pi$                                &$10.225\pi$                                     &$11.961\pi$\\
  {$10$dB}           &$7.861\pi$                                &$10.163\pi$                                     &$11.720\pi$\\
  {$15$dB}           &$8.045\pi$                                &$9.901\pi$                                     &$11.628\pi$\\
  {$20$dB}           &$8.000\pi$                                &$10.004\pi$                                     &$11.495\pi$\\
  \bottomrule
  \label{Table}
\end{tabular}}
\end{center}
\vspace{-1.0cm}
\end{table}
\begin{figure}[t]
\setlength{\abovecaptionskip}{-0.1cm}   %è°æŽåŸçæ é¢äžåŸè·çŠ»
\setlength{\belowcaptionskip}{-0.4cm}   %è°æŽåŸçæ é¢äžäžæè·çŠ»
\centering
\subfigure[]{
\includegraphics[scale=0.485]{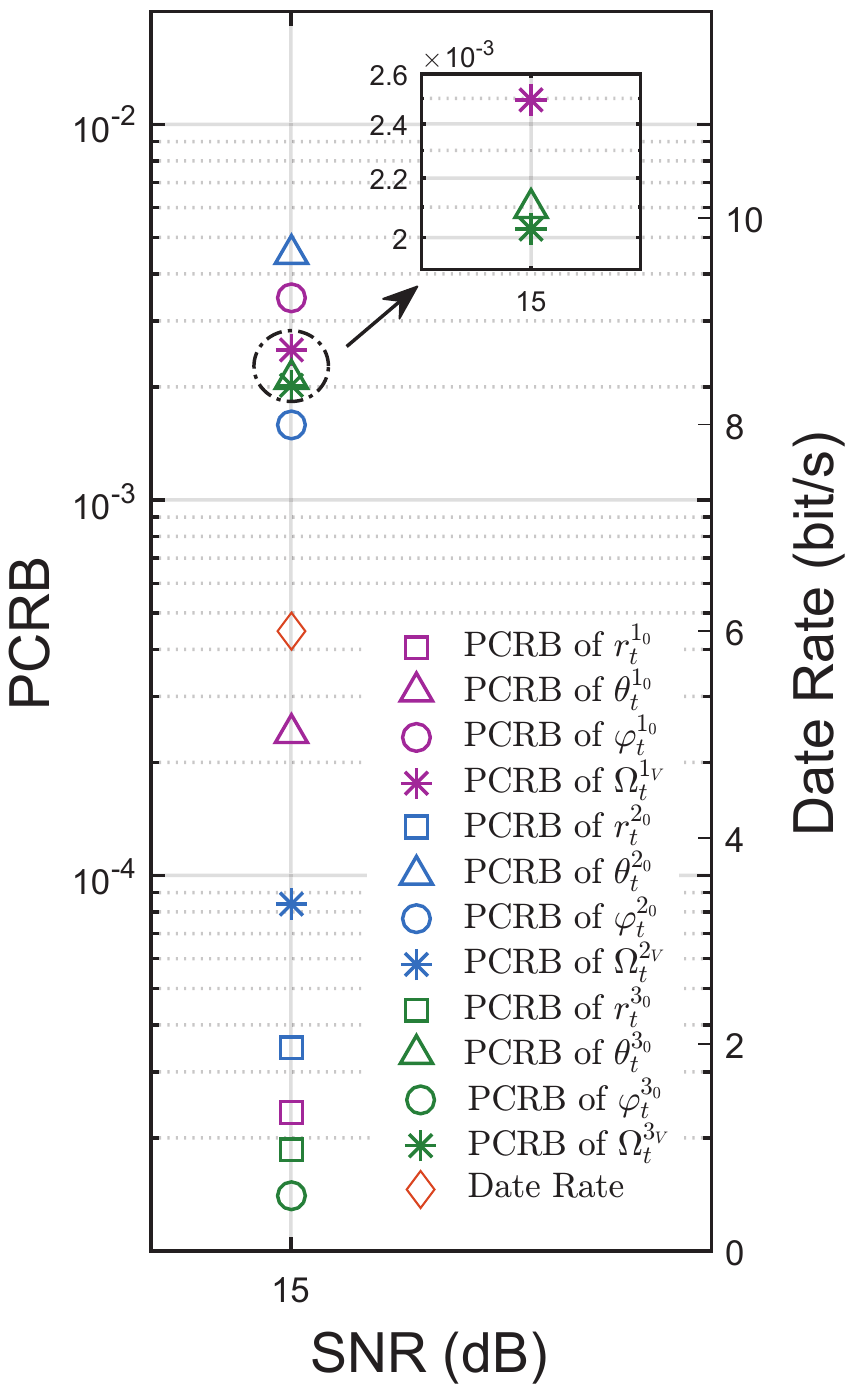}
}
\subfigure[]{
\includegraphics[scale=0.485]{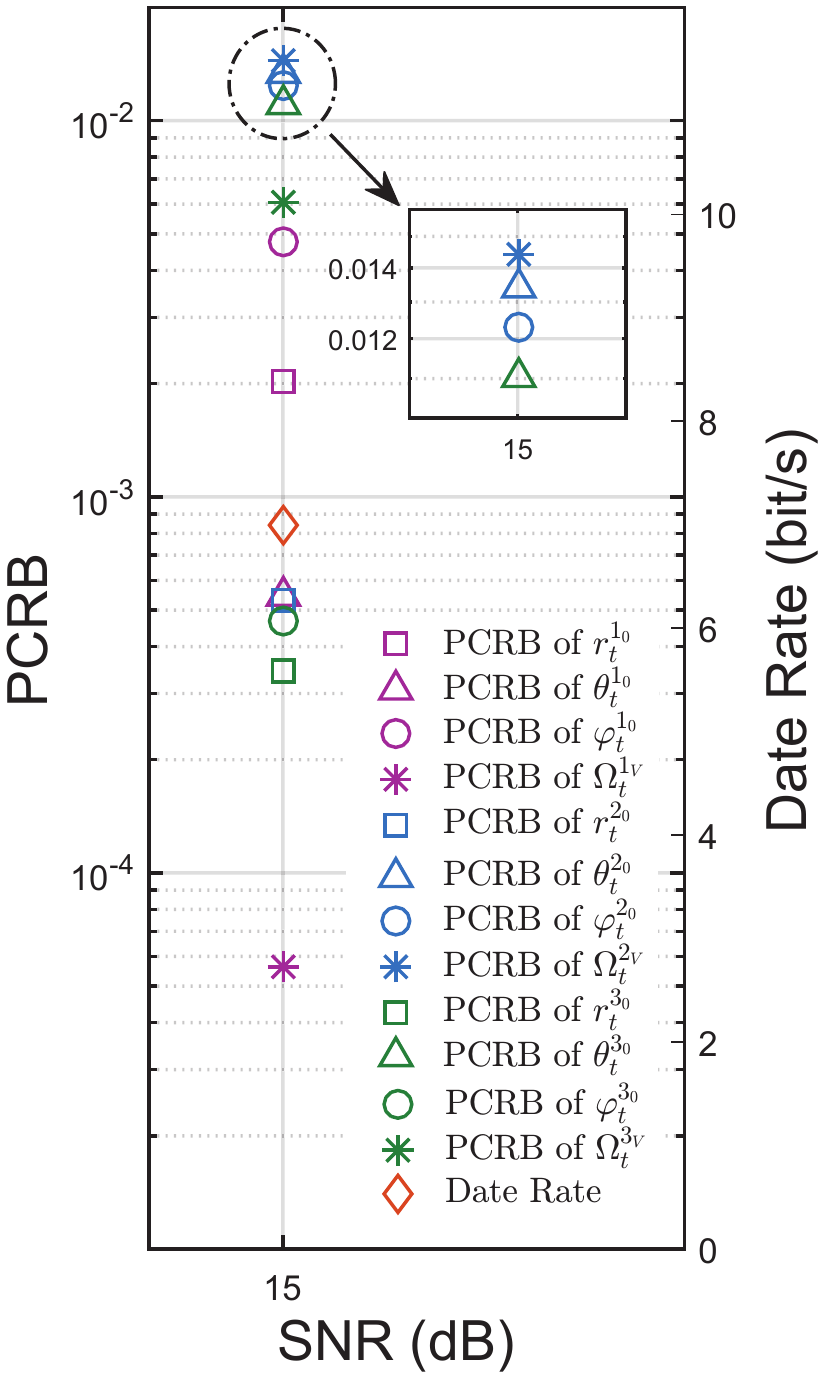}
}
\caption{The minimum PCRBs under the different data rate constraints. (a) The required data rate $= 6$ bit/s. (b) The required data rate $= 7$ bit/s.}
\label{Fig9}
\end{figure}
Fig.\ref{Fig7} shows the time-frequency distributions of the vertices of $Q$ targets. It can be seen the theoretical curves  almost coincide with the time-frequency simulation results, proving the effectiveness of the rotational Doppler shift model derived in Section III.B. Thereafter, the estimated rotation velocities are shown in Table I.  The estimated measures  approach the actual velocities as the the SNR increases. Meanwhile, noise has little influence on the  rotation velocity detection when using the STFT method.
%In Fig.\ref{Fig7}, we suppose that the rotation velocities of $Q$ targets are $\Omega^{1}_{t}=8\pi$, $\Omega^{2}_{t}=10\pi$ and $\Omega^{3}_{t}=11.5\pi$. Then, the time-frequency distributions of the vertices of $Q$ targets only consisting of the rotational Doppler shift are shown in Fig.\ref{Fig7}. It can be seen from the figure that the curves of the theoretical rotational Doppler shift of the vertices almost coincides with the time-frequency simulation results, which proves the effectiveness of the rotational Doppler shift model derived in Section III.B. After that, by using the proposed method, the estimated rotation velocities of $Q$ targets are shown in Table I. As we can see in Table I, estimated velocities of $Q$ targets approach the actual velocities with the increase of the SNR. Meanwhile, the noises have little influence on the OAM-based rotation velocity detection when using the STFT method.

In Fig.\ref{addfig}, we compare the MSEs and PCRBs of the estimates $\{(\hat{r}^{q_0}_t,\hat{\theta}^{q_0}_t,\hat{\varphi}^{q_0}_t,\hat{\Omega}^{q_V}_t)|q=1,2,3\}$ under the condition of average power distribution. The MSE is defined as $\mathbb{E}\{{(\hat{x}-x)}^2\}$, where $\hat{x}$ denotes the estimate of $x$. As the SNR increases, the MSEs of the estimates  decrease and approach the PCRBs gradually. At high SNRs, the MSEs of the estimates $\{(\hat{r}^{q_0}_t,\hat{\theta}^{q_0}_t,\hat{\varphi}^{q_0}_t,\hat{\Omega}^{q_V}_t)|q=1,2,3\}$ are very close to their PCRBs, proving the effectiveness of the proposed OAM-based rotating target imaging method.

In Fig.\ref{Fig8}, we show the PCRB and data rate of the radar-centric joint OAM RadCom system under QPSK modulation. %It is obvious to see from the figure that the imaging and communication performances of the joint OAM RadCom system are greatly improved with the increase of the transmit SNR. More importantly,
Comparing the two figures, we can see that the weight distribution of integrated OAM beams directly affects the PCRB and data rate of the system, especially the PCRB, which indicates the necessity of the weight optimization for integrated OAM beams.

In Fig.\ref{Fig9}, we assume that the number of intermediate points is $\mathfrak{N}=10$, the transmit SNR is $15$dB, and the data rates required by the specific target  are $6$ bit/s in in Fig.\ref{Fig9} (a)  and $7$ bit/s in Fig.\ref{Fig9} (b).
%Then, the minimum PCRBs of the radar-centric joint OAM RadCom system under the different data rate constraints are shown in Fig.\ref{Fig9}.
It can be seen from the figure that under different data rate constraints, the PCRBs of the system have different optimization results. More importantly, comparing Fig.\ref{Fig8} and Fig.\ref{Fig9}, we can see that by using the proposed optimization method, the PCRBs of the joint OAM RadCom system are greatly improved while the data rates of the specific target are guaranteed,  realizeing the best performance tradeoff of the system.

\section{Conclusions}

In this paper, we propose a novel UCA-based radar-centric joint OAM RadCom scheme including the OAM-based 3-D position estimation, rotation velocity detection and specific target communication. In terms of the OAM-based position estimation, we propose a 3-D super-resolution position estimation method based on the MUSIC algorithm. By using the super-resolution MUSIC algorithm for OAM echo signals in the frequency domain and OAM mode domain, the proposed method breaks through the limitation of the elevation resolution of the existing OAM-based imaging method and is capable of the 3-D super-resolution position estimation of multiple targets. Moreover, we propose a rotation velocity detection method based on the STFT algorithm, which takes full advantage of the rotational Doppler characteristics of OAM echo signals to estimate the rotation velocities of multiple targets simultaneously, providing a basis for distinguishing the type of targets. Thereafter, we analyze the PCRB and data rate of the radar-centric joint OAM RadCom system, and then optimize the transmitted integrated OAM beams by applying an exhaustive search method to achieve an optimal performance tradeoff. Both mathematical analysis and simulation results show that  the proposed radar-centric joint OAM RadCom scheme can accurately estimate the 3-D position and rotation velocity of multiple targets, while ensuring the data rate of the specific target.
%realizing spectrum sharing of wireless communication and radar sensing services.
%the 3-D position of all the nodes can be accurately estimated

\begin{appendices}
\section{ }
The Fisher information matrix $\mathbf{J}$ with respect to $\boldsymbol{\vartheta}$ can be expressed as
\begin{align}\label{Fisher2}
\mathbf{J}=&\mathbb{E}\left\{\!\left[\frac{\partial}{\partial\boldsymbol{\vartheta}}\ln f(\mathbf{E}_R,\boldsymbol{\vartheta})\right]\!\!\!\left[\frac{\partial}{\partial\boldsymbol{\vartheta}}\ln f(\mathbf{E}_R,\boldsymbol{\vartheta})\right]^\mathrm{T}\!\right\}\nonumber\\
=&\mathbb{E}\left\{\!\left[\frac{\partial}{\partial\boldsymbol{\vartheta}}\ln f(\mathbf{E}_R|\boldsymbol{\vartheta})\right]\!\!\!\left[\frac{\partial}{\partial\boldsymbol{\vartheta}}\ln f(\mathbf{E}_R|\boldsymbol{\vartheta})\right]^\mathrm{T}\!\right\}\nonumber\\
&+\mathbb{E}\left\{\!\left[\frac{\partial}{\partial\boldsymbol{\vartheta}}\ln f(\boldsymbol{\vartheta})\right]\!\!\!\left[\frac{\partial}{\partial\boldsymbol{\vartheta}}\ln f(\boldsymbol{\vartheta})\right]^\mathrm{T}\!\right\}\nonumber\\
&+\mathbb{E}\left\{\!\left[\frac{\partial}{\partial\boldsymbol{\vartheta}}\ln f(\mathbf{E}_R|\boldsymbol{\vartheta})\right]\!\!\!\left[\frac{\partial}{\partial\boldsymbol{\vartheta}}\ln f(\boldsymbol{\vartheta})\right]^\mathrm{T}\!\right\}\nonumber\\
&+\mathbb{E}\left\{\!\left[\frac{\partial}{\partial\boldsymbol{\vartheta}}\ln f(\boldsymbol{\vartheta})\right]\!\!\!\left[\frac{\partial}{\partial\boldsymbol{\vartheta}}\ln f(\mathbf{E}_R|\boldsymbol{\vartheta})\right]^\mathrm{T}\!\right\}\nonumber\\
=&\mathbf{J}_A+\mathbf{J}_B+\mathbf{J}_C+\mathbf{J}_D,
\end{align}
where $\ln f(\mathbf{E}_R,\boldsymbol{\vartheta})=\ln f(\mathbf{E}_R|\boldsymbol{\vartheta})+\ln f(\boldsymbol{\vartheta})$ is the joint probability density of $(\mathbf{E}_R,\boldsymbol{\vartheta})$, $\ln f(\mathbf{E}_R|\boldsymbol{\vartheta})$ is the conditional probability density of $(\mathbf{E}_R,\boldsymbol{\vartheta})$, and $\ln f(\boldsymbol{\vartheta})$ is the marginal probability density of $\boldsymbol{\vartheta}$. Due to $\int_{(\mathbf{E}_R)}f(\mathbf{E}_R|\boldsymbol{\vartheta}) \mathrm{d} \mathbf{E}_R =1$,
\begin{align}\label{Derivative}
&\frac{\partial}{\partial \vartheta_i}\int_{(\mathbf{E}_R)}f(\mathbf{E}_R|\boldsymbol{\vartheta}) \mathrm{d} \mathbf{E}_R =\int_{(\mathbf{E}_R)}\frac{\partial}{\partial \vartheta_i}f(\mathbf{E}_R|\boldsymbol{\vartheta}) \mathrm{d} \mathbf{E}_R\nonumber\\
&=\int_{(\mathbf{E}_R)}f(\mathbf{E}_R|\boldsymbol{\vartheta})\frac{\partial}{\partial \vartheta_i}\ln f(\mathbf{E}_R|\boldsymbol{\vartheta}) \mathrm{d} \mathbf{E}_R=0.
\end{align}
Thus, the $i$-th row and $j$-th column element $[\mathbf{J}_C]_{ij}$ of $\mathbf{J}_C$ and the $j$-th row and $i$-th column element $[\mathbf{J}_D]_{ji}$ of $\mathbf{J}_D$ are derived as
\begin{align}\label{Derivative2}
\!\![\mathbf{J}&_C]_{ij}=[\mathbf{J}_D]_{ji}=\mathbb{E}\left\{\!\left[\frac{\partial}{\partial \vartheta_i}\ln f(\mathbf{E}_R|\boldsymbol{\vartheta})\right]\!\!\!\left[\frac{\partial}{\partial \vartheta_j}\ln f(\boldsymbol{\vartheta})\right]\!\right\}\nonumber\\
=&\int_{(\boldsymbol{\vartheta})}\!\int_{(\mathbf{E}_R)}\!\!f(\mathbf{E}_R|\boldsymbol{\vartheta})\frac{\partial}{\partial \vartheta_i}\!\ln \! f(\mathbf{E}_R|\boldsymbol{\vartheta})f(\boldsymbol{\vartheta})\frac{\partial}{\partial \vartheta_j}\ln\!f(\boldsymbol{\vartheta}) \mathrm{d}\mathbf{E}_R\mathrm{d}\boldsymbol{\vartheta}\nonumber\\
=&0,
\end{align}
that is, $\mathbf{J}_C=\mathbf{0}$ and $\mathbf{J}_D=\mathbf{0}$. After that, the Fisher information matrix $\mathbf{J}$ in \eqref{Fisher2} is simplified to
\begin{align}\label{Fisher3}
\mathbf{J}=&\mathbb{E}\left\{\!\left[\frac{\partial}{\partial\boldsymbol{\vartheta}}\ln f(\mathbf{E}_R|\boldsymbol{\vartheta})\right]\!\!\!\left[\frac{\partial}{\partial\boldsymbol{\vartheta}}\ln f(\mathbf{E}_R|\boldsymbol{\vartheta})\right]^\mathrm{T}\!\right\}\nonumber\\
&+\mathbb{E}\left\{\!\left[\frac{\partial}{\partial\boldsymbol{\vartheta}}\ln f(\boldsymbol{\vartheta})\right]\!\!\!\left[\frac{\partial}{\partial\boldsymbol{\vartheta}}\ln f(\boldsymbol{\vartheta})\right]^\mathrm{T}\!\right\}\nonumber\\
=&\mathbf{J}_A+\mathbf{J}_B,
\end{align}
where $\mathbf{J}_A$ represents the information obtained from OAM echo signals, $\mathbf{J}_B$ represents the priori information.

Considering that $\{n(\ell_u,k_w)$$|$$u$$=$$1,$$2,$$\cdots,$$U,$$w$$=$$1,$$2,$$\cdots,$$W\}$ are independent AWGNs with zero mean and variance $\xi^2$, the conditional probability density $\ln f(\mathbf{E}_R|\boldsymbol{\vartheta})$ can be written as
\begin{align}\label{CPD}
\ln f(\mathbf{E}_R|\boldsymbol{\vartheta})=&-UW\ln\sqrt{2\pi}\xi\nonumber\\
&-\!\frac{1}{2\xi^2}\sum_{u=1}^{U}\!\sum_{w=1}^{W}\!\big|E_R(\ell_u,k_w)\!-\!p(\ell_u,k_w,\boldsymbol{\vartheta})\big|^2.
\end{align}
Then, the $i$-th row and $j$-th column element $[\mathbf{J}_A]_{ij}$ of $\mathbf{J}_A$ can be expressed as
\begin{align}\label{FisherA}
[\mathbf{J}_A]_{ij}=&\mathbb{E}\left\{\!\left[\frac{\partial}{\partial \vartheta_i}\ln f(\mathbf{E}_R|\boldsymbol{\vartheta})\right]\!\!\!\left[\frac{\partial}{\partial\vartheta_j}\ln f(\mathbf{E}_R|\boldsymbol{\vartheta})\right]\!\right\}\nonumber\\
=&-\mathbb{E}\left\{\!\frac{\partial}{\partial \vartheta_i}\left[\frac{\partial}{\partial\vartheta_j}\ln f(\mathbf{E}_R|\boldsymbol{\vartheta})\right]\!\right\}\nonumber\\
=&\mathbb{E}\left\{\!\frac{1}{\xi^2}\sum_{u=1}^{U}\!\sum_{w=1}^{W}\textrm{Re}\bigg[\frac{\partial p(\ell_u,k_w,\boldsymbol{\vartheta})}{\partial \vartheta_i}\frac{\partial p(\ell_u,k_w,\boldsymbol{\vartheta})}{\partial \vartheta_j}\!\bigg]\right\}\nonumber\\
&-\mathbb{E}\left\{\!\frac{1}{\xi^2}\sum_{u=1}^{U}\!\sum_{w=1}^{W}\textrm{Re}\bigg[n(\ell_u,k_w)\frac{\partial^2 p(\ell_u,k_w,\boldsymbol{\vartheta})}{\partial \vartheta_i\partial \vartheta_j}\!\bigg]\right\}\nonumber\\
=&\frac{1}{\xi^2}\sum_{u=1}^{U}\!\sum_{w=1}^{W}\textrm{Re}\bigg[\frac{\partial p(\ell_u,k_w,\boldsymbol{\vartheta})}{\partial \vartheta_i}\frac{\partial p(\ell_u,k_w,\boldsymbol{\vartheta})}{\partial \vartheta_j}\bigg].
\end{align}

Supposing that the key state parameters to be estimated in $\boldsymbol{\vartheta}$ satisfy uniform distribution, i.e., $r^{q_0}_t\sim U(0,r_m)$, $\theta^{q_0}_t\sim U(0,\pi)$, $\varphi^{q_0}_t\sim U(0,2\pi)$ and $\Omega^{q_{_V}}_t\sim U(0,\Omega_m)$, where $r_m$ and $\Omega_m$ are the longest distance and maximum rotation velocity that can be detected by the OAM-based target imaging, $q=,1,2,\cdots,Q$. Then, the marginal probability density $\ln f(\boldsymbol{\vartheta})$ can be written as
\begin{align}\label{PD}
\ln f(\boldsymbol{\vartheta})&=-Q\ln2\pi^2r_m\Omega_m.
\end{align}
Thus, the $i$-th row and $j$-th column element $[\mathbf{J}_B]_{ij}$ of $\mathbf{J}_B$ is written as
\begin{align}\label{FisherB}
[\mathbf{J}_B]_{ij}=&\mathbb{E}\left\{\!\left[\frac{\partial}{\partial \vartheta_i}\ln f(\boldsymbol{\vartheta})\right]\!\!\!\left[\frac{\partial}{\partial\vartheta_j}\ln f(\boldsymbol{\vartheta})\right]\!\right\}\nonumber\\
=&-\mathbb{E}\left\{\!\frac{\partial}{\partial \vartheta_i}\left[\frac{\partial}{\partial\vartheta_j}\ln f(\boldsymbol{\vartheta})\right]\!\right\}=0,
\end{align}
i.e., $\mathbf{J}_B=\mathbf{0}$. Finally, the Fisher information matrix $\mathbf{J}$ is derived as \eqref{Fisher}.
\end{appendices}

\bibliographystyle{IEEEtran}
\bibliography{IEEEabrv,ref}

\begin{thebibliography}{10}
\providecommand{\url}[1]{#1}
\csname url@samestyle\endcsname
\providecommand{\newblock}{\relax}
\providecommand{\bibinfo}[2]{#2}
\providecommand{\BIBentrySTDinterwordspacing}{\spaceskip=0pt\relax}
\providecommand{\BIBentryALTinterwordstretchfactor}{4}
\providecommand{\BIBentryALTinterwordspacing}{\spaceskip=\fontdimen2\font plus
\BIBentryALTinterwordstretchfactor\fontdimen3\font minus
  \fontdimen4\font\relax}
\providecommand{\BIBforeignlanguage}[2]{{%
\expandafter\ifx\csname l@#1\endcsname\relax
\typeout{** WARNING: IEEEtran.bst: No hyphenation pattern has been}%
\typeout{** loaded for the language `#1'. Using the pattern for}%
\typeout{** the default language instead.}%
\else
\language=\csname l@#1\endcsname
\fi
#2}}
\providecommand{\BIBdecl}{\relax}
\BIBdecl

\bibitem{Zhang20196G}
Z.~Zhang, Y.~Xiao, Z.~Ma, M.~Xiao, Z.~Ding, X.~Lei, G.~K. Karagiannidis, and
  P.~Fan, ``{6G} wireless networks: Vision, requirements, architecture, and key
  technologies,'' \emph{IEEE Veh. Technol. Mag.}, vol.~14, no.~3, pp. 28--41,
  Sept. 2019.

\bibitem{Long2021A}
W.-X. Long, R.~Chen, M.~Moretti, W.~Zhang, and J.~Li, ``A promising technology
  for {6G} wireless networks: Intelligent reflecting surface,'' \emph{J.
  Commun. Inf. Networks}, vol.~6, no.~1, pp. 1--16, Mar. 2021.

\bibitem{WRC}
(2018, Sept.) World radiocommunication conference ({WRC}). [Online]. Available:
  \url{https://www.itu.int/en/ITU-R/conferences/wrc/ Pages/default.aspx}.

\bibitem{zhang2021overview}
J.~A. Zhang, F.~Liu, C.~Masouros, R.~W. Heath, Z.~Feng, L.~Zheng, and
  A.~Petropulu, ``An overview of signal processing techniques for joint
  communication and radar sensing,'' \emph{IEEE J. Sel. Topics Signal
  Process.}, vol.~15, no.~6, pp. 1295--1315, Nov. 2021.

\bibitem{Zheng2019Radar}
L.~Zheng, M.~Lops, Y.~C. Eldar, and X.~Wang, ``Radar and communication
  coexistence: An overview: A review of recent methods,'' \emph{IEEE Signal
  Process. Mag.}, vol.~36, no.~5, pp. 85--99, Sept. 2019.

\bibitem{Feng2020Joint}
Z.~Feng, Z.~Fang, Z.~Wei, X.~Chen, Z.~Quan, and D.~Ji, ``Joint radar and
  communication: {A} survey,'' \emph{China Commun.}, vol.~17, no.~1, pp. 1--27,
  Jan. 2020.

\bibitem{Mealey1963A}
M.~R. M., ``A method for calculating error probabilities in a radar
  communication system,'' \emph{IEEE Trans. Space Electron. Telemetry}, vol.~9,
  no.~2, pp. 37--42, Jun. 1963.

\bibitem{Quan2014Radar}
S.~Quan, W.~Qian, J.~Guq, and V.~Zhang, ``Radar-communication integration: {An}
  overview,'' in \emph{7th IEEE/Int'l. Conf. Advanced Infocomm Technology},
  2014, pp. 98--103.

\bibitem{Liu2018Toward}
F.~Liu, L.~Zhou, C.~Masouros, A.~Li, W.~Luo, and A.~Petropulu, ``Toward
  dual-functional radar-communication systems: Optimal waveform design,''
  \emph{IEEE Trans. Signal Process.}, vol.~66, no.~16, pp. 4264--4279, Aug.
  2018.

\bibitem{Zheng2018On}
Y.~L. Sit, B.~Nuss, and T.~Zwick, ``On mutual interference cancellation in a
  {MIMO OFDM} multiuser radar-communication network,'' \emph{IEEE Trans. Veh.
  Technol.}, vol.~67, no.~4, pp. 3339--3348, Apr. 2018.

\bibitem{rihan2018non}
M.~Rihan and L.~Huang, ``Non-orthogonal multiple access based cooperative
  spectrum sharing between {MIMO} radar and {MIMO} communication systems,''
  \emph{Digit. Signal Process.}, vol.~83, pp. 107--117, Dec. 2018.

\bibitem{Ahmed2019Distributed}
A.~Ahmed, Y.~D. Zhang, and B.~Himed, ``Distributed dual-function
  radar-communication {MIMO} system with optimized resource allocation,'' in
  \emph{2019 IEEE Radar Conference (RadarConf)}, 2019, pp. 1--5.

\bibitem{Keskin2021MIMO}
M.~F. Keskin, H.~Wymeersch, and V.~Koivunen, ``{MIMO-OFDM} joint
  radar-communications: Is {ICI} friend or foe?'' \emph{IEEE J. Sel. Topics
  Signal Process.}, vol.~15, no.~6, pp. 1393--1408, Nov. 2021.

\bibitem{Zhou2021Performance}
X.~Zhou, L.~Tang, Y.~Bai, and Y.-C. Liang, ``Performance analysis and waveform
  optimization of integrated {FD-MIMO} radar-communication systems,''
  \emph{IEEE Trans. Wireless Commun.}, vol.~20, no.~11, pp. 7490--7502, Nov.
  2021.

\bibitem{Allen1992Orbital}
L.~Allen, M.~W. Beijersbergen, R.~J. Spreeuw, and J.~P. Woerdman, ``Orbital
  angular momentum of light and the transformation of {Laguerre-Gaussian} laser
  modes,'' \emph{Phys. Rev. A: At. Mol. Opt. Phys.}, vol.~45, no.~11, pp.
  8185--8189, June 1992.

\bibitem{Guo2013Electromagnetic}
G.~Guo, W.~Hu, and X.~Du, ``Electromagnetic vortex based radar target
  imaging,'' \emph{J. Nat. Univ. Defense Technol.}, vol.~35, no.~6, pp. 71--76,
  Dec. 2013.

\bibitem{lin2016improved}
M.~Lin, Y.~Gao, P.~Liu, and J.~Liu, ``Improved {OAM}-based radar targets
  detection using uniform concentric circular arrays,'' \emph{Int. J. Antenn.
  Propag.}, vol. 2016, no.~6, pp. 1--8, Jan. 2016.

\bibitem{Chen2018OAMradar}
R.~Chen, W.-X. Long, Y.~Gao, and J.~Li, ``Orbital angular momentum-based
  two-dimensional super-resolution targets imaging,'' in \emph{Proc. IEEE
  Global Conf. Signal Inf. Process.}, 2018, pp. 1--4.

\bibitem{Liu2020Microwave}
H.~Liu, K.~Liu, Y.~Cheng, and H.~Wang, ``Microwave vortex imaging based on dual
  coupled {OAM} beams,'' \emph{IEEE Sens. J}, vol.~20, no.~2, pp. 806--815,
  Jan. 2020.

\bibitem{wang2021Object}
J.~Wang, K.~Liu, H.~Liu, K.~Cao, Y.~Cheng, and H.~Wang, ``{3-D} object imaging
  method with electromagnetic vortex,'' \emph{IEEE Trans. Geosci. Remote
  Sens.}, vol.~60, pp. 1--12, Apr. 2022.

\bibitem{Yuan2021Resolution}
H.~Yuan, Y.~Chen, Y.~Luo, J.~Liang, and Z.~Wang, ``A resolution-improved
  imaging algorithm based on uniform circular array,'' \emph{IEEE Antennas
  Wireless Propag. Lett.}, pp. 1--1, 2021.

\bibitem{Lavery2013Detection}
M.~P. Lavery, F.~C. Speirits, S.~M. Barnett, and M.~J. Padgett, ``Detection of
  a spinning object using light¡¯s orbital angular momentum,''
  \emph{Science}, vol. 341, no. 6145, pp. 537--540, Sept. 2013.

\bibitem{Courtial1998Measurement}
J.~Courtial, K.~Dholakia, D.~Robertson, L.~Allen, and M.~Padgett, ``Measurement
  of the rotational frequency shift imparted to a rotating light beam
  possessing orbital angular momentum,'' \emph{Phys. Rev. Lett.}, vol.~80,
  no.~15, pp. 3217--3219, Apr. 1998.

\bibitem{Zhao2016Measurement}
M.~Zhao, X.~Gao, M.~Xie, W.~Zhai, W.~Xu, S.~Huang, and W.~Gu, ``Measurement of
  the rotational doppler frequency shift of a spinning object using a radio
  frequency orbital angular momentum beam,'' \emph{Opt. Lett.}, vol.~41,
  no.~11, pp. 2549--2552, Jun. 2016.

\bibitem{Liu2017Microwave}
K.~Liu, X.~Li, Y.~Gao, H.~Wang, and Y.~Cheng, ``Microwave imaging of spinning
  object using orbital angular momentum,'' \emph{J. Appl. Phys.}, vol. 122,
  no.~12, p. 124903, Sept. 2017.

\bibitem{Cole2010Missile}
C.~E. Cole, ``Missile communication links,'' \emph{Johns Hopkins APL Technical
  Digest}, vol.~28, no.~4, pp. 324--330, 2010.

\bibitem{Chen2020Orbital}
R.~Chen, H.~Zhou, M.~Moretti, X.~Wang, and J.~Li, ``Orbital angular momentum
  waves: Generation, detection, and emerging applications,'' \emph{IEEE Commun.
  Surveys Tuts.}, vol.~22, no.~2, pp. 840--868, Nov. 2020.

\bibitem{Tamburini2012Encoding}
F.~Tamburini, E.~Mari, A.~Sponselli, B.~Thid\'e, A.~Bianchini, and F.~Romanato,
  ``Encoding many channels in the same frequency through radio vorticity: first
  experimental test,'' \emph{New J. Phys.}, vol.~14, no.~3, p. 033001, Mar.
  2012.

\bibitem{Mahmouli20134}
F.~E. Mahmouli and S.~D. Walker, ``{4-Gbps} uncompressed video transmission
  over a {60-GHz} orbital angular momentum wireless channel,'' \emph{IEEE
  Wireless Commun. Lett.}, vol.~2, no.~2, pp. 223--226, Apr. 2013.

\bibitem{Yan2014High}
Y.~Yan, G.~Xie, M.~P.~J. Lavery, and et~al., ``High-capacity millimetre-wave
  communications with orbital angular momentum multiplexing,'' \emph{Nature
  Commun.}, vol.~5, p. 4876, Sept. 2014.

\bibitem{Ren2017Line}
Y.~Ren, L.~Li, G.~Xie, and et~al., ``Line-of-sight millimeter-wave
  communications using orbital angular momentum multiplexing combined with
  conventional spatial multiplexing,'' \emph{{IEEE} Trans. Wireless Commun.},
  vol.~16, no.~5, pp. 3151--3161, May 2017.

\bibitem{Chen2018A}
R.~Chen, W.~Yang, H.~Xu, and J.~Li, ``A 2-{D} {FFT}-based transceiver
  architecture for {OAM-OFDM} systems with {UCA} antennas,'' \emph{{IEEE}
  Trans. Veh. Technol.}, vol.~67, no.~6, pp. 5481--5485, June 2018.

\bibitem{Chen2020Multi}
R.~Chen, W.-X. Long, X.~Wang, and L.~Jiandong, ``Multi-mode oam radio waves:
  {G}eneration, angle of arrival estimation and reception with {UCAs},''
  \emph{IEEE Trans. Wireless Commun.}, vol.~19, no.~10, pp. 6932--6947, Oct.
  2020.

\bibitem{Long2021Joint}
W.-X. Long, R.~Chen, M.~Moretti, J.~Xiong, and J.~Li, ``Joint spatial division
  and coaxial multiplexing for downlink multi-user {OAM} wireless backhaul,''
  \emph{IEEE Trans. Broadcast.}, vol.~67, no.~4, pp. 879--893, Dec. 2021.

\bibitem{Long2021AoA}
W.-X. Long, R.~Chen, M.~Moretti, and J.~Li, ``{AoA} estimation for {OAM}
  communication systems with mode-frequency multi-time {ESPRIT} method,''
  \emph{IEEE Trans. Veh. Technol.}, vol.~70, no.~5, pp. 5094--5098, May 2021.

\bibitem{Yagi2021200}
Y.~Yagi, H.~Sasaki, T.~Yamada, and D.~Lee, ``200 {Gb/s} wireless transmission
  using dual-polarized {OAM-MIMO} multiplexing with uniform circular array on
  28 {GHz} band,'' \emph{IEEE Antennas Wireless Propag. Lett.}, vol.~20, no.~5,
  pp. 833--837, May 2021.

\bibitem{Liu2016Generation}
K.~Liu, Y.~Cheng, X.~Li, Y.~Qin, H.~Wang, and Y.~Jiang, ``Generation of orbital
  angular momentum beams for electromagnetic vortex imaging,'' \emph{IEEE
  Antennas and Wirel. Propag. Lett.}, vol.~15, pp. 1873--1876, Mar. 2016.

\bibitem{Liu2015Orbital}
K.~{Liu}, Y.~{Cheng}, Z.~{Yang}, H.~{Wang}, Y.~{Qin}, and X.~{Li},
  ``Orbital-angular-momentum-based electromagnetic vortex imaging,'' \emph{IEEE
  Antennas Wireless Propag. Lett.}, vol.~14, pp. 711--714, Dec. 2015.

\bibitem{Bekar2021Joint}
M.~Bekar, C.~J. Baker, E.~G. Hoare, and M.~Gashinova, ``Joint {MIMO} radar and
  communication system using a {PSK-LFM} waveform with {TDM} and {CDM}
  approaches,'' \emph{IEEE Sens. J}, vol.~21, no.~5, pp. 6115--6124, Mar. 2021.

\bibitem{Schmidt1896Multiple}
R.~Schmidt, ``Multiple emitter location and signal parameter estimation,''
  \emph{IEEE Trans. Antennas Propag.}, vol.~34, no.~3, pp. 276--280, Mar. 1986.

\bibitem{Gao2010Doppler}
H.~Gao, L.~Xie, S.~Wen, and Y.~Kuang, ``Micro-{Doppler} signature extraction
  from ballistic target with micro-motions,'' \emph{IEEE Trans. Aerosp.
  Electron. Syst.}, vol.~46, no.~4, pp. 1969--1982, Oct. 2010.

\bibitem{Lei2012Micromotion}
P.~Lei, J.~Sun, J.~Wang, and W.~Hong, ``Micromotion parameter estimation of
  free rigid targets based on radar micro-doppler,'' \emph{IEEE Trans. Geosci.
  Remote Sens.}, vol.~50, no.~10, pp. 3776--3786, Oct. 2012.

\bibitem{Wang2021Detection}
Y.~Wang, K.~Liu, H.~Liu, J.~Wang, and Y.~Cheng, ``Detection of rotational
  object in arbitrary position using vortex electromagnetic waves,'' \emph{IEEE
  Sensors J.}, vol.~21, no.~4, pp. 4989--4994, Feb. 2021.

\bibitem{Chen1998Joint}
V.~Chen and S.~Qian, ``Joint time-frequency transform for radar range-{Doppler}
  imaging,'' \emph{IEEE Trans. Aerosp. Electron. Syst.}, vol.~34, no.~2, pp.
  486--499, Apr. 1998.

\bibitem{Tichavsky1998Posterior}
P.~Tichavsky, C.~Muravchik, and A.~Nehorai, ``Posterior cram\'{e}r-rao bounds
  for discrete-time nonlinear filtering,'' \emph{IEEE Trans. Signal Process.},
  vol.~46, no.~5, pp. 1386--1396, May 1998.

\bibitem{NAYAK202071Single}
S.~Nayak, \emph{Fundamentals of Optimization Techniques with Algorithms}.\hskip
  1em plus 0.5em minus 0.4em\relax Academic Press, 2020.

\end{thebibliography}

\end{document}